\documentclass [10pt,twocolumn]{IEEEtran}
\usepackage{algorithm,amsbsy,amsmath,amssymb,epsfig,bbm,mathrsfs,multirow,amsthm}
\usepackage[T1]{fontenc}
\usepackage[latin9]{inputenc}
\usepackage{amsmath}
\usepackage{fixltx2e}
\usepackage{algorithm}
\usepackage{array,multirow,graphicx}
\usepackage{color}
\usepackage{cite}
\usepackage{float}
\usepackage{makecell}
\usepackage[x11names,dvipsnames,table]{xcolor} %for use in color links
\usepackage{colortbl}
\usepackage{multirow}
\usepackage{adjustbox}
\usepackage{subcaption} 
\usepackage{lipsum}
\definecolor{mygray}{gray}{0.6}
\definecolor{myblue}{rgb}{0.8,0.85,1} 
\usepackage{array}
\newcolumntype{L}[1]{>{\raggedright\let\newline\\\arraybackslash\hspace{0pt}}m{#1}}
\newcolumntype{C}[1]{>{\centering\let\newline\\\arraybackslash\hspace{0pt}}m{#1}}
\newcolumntype{R}[1]{>{\raggedleft\let\newline\\\arraybackslash\hspace{0pt}}m{#1}}

\usepackage{array, tabularx, boldline}
\usepackage{eurosym}
\usepackage{amstext} % for \text
\DeclareRobustCommand{\officialeuro}{%
\ifmmode\expandafter\text\fi
{\fontencoding{U}\fontfamily{eurosym}\selectfont e}}

\usepackage{booktabs}
\usepackage{multirow}

\usepackage{array, tabularx, boldline}
\usepackage{graphicx}
\usepackage{cellspace}
\usepackage{algpseudocode}
\theoremstyle{plain}

\theoremstyle{definition}

\providecommand{\definitionname}{Definition}
\providecommand{\theoremname}{Theorem}

\begin{document}

\title{Federated Learning in Mobile Edge Networks: A Comprehensive Survey}
\author{
Wei Yang Bryan Lim\thanks{W. Y. B. Lim is with Alibaba Group and the Alibaba-NTU Joint Research Institute, Nanyang Technological University, Singapore. Email: limw0201@e.ntu.edu.sg.}, 
Nguyen Cong Luong\thanks{N. C.~Luong is with Faculty of Information Technology, PHENIKAA University, Hanoi 12116, Vietnam, and is with PHENIKAA Research and Technology Institute (PRATI), A\&A Green Phoenix
Group JSC, No.167 Hoang Ngan, Trung Hoa, Cau Giay, Hanoi 11313, Vietnam. Email: luong.nguyencong@phenikaa-uni.edu.vn.}, 
Dinh Thai Hoang\thanks{D.~T.~Hoang is with the School of Electrical and Data Engineering, University of Technology Sydney, Australia. E-mail: hoang.dinh@uts.edu.au.},
Yutao Jiao\thanks{Y. Jiao,  D.~Niyato and C. Miao  are with School of Computer Science and Engineering, Nanyang Technological University, Singapore. E-mails:  
yjiao001@e.ntu.edu.sg, dniyato@ntu.edu.sg, ascymiao@ntu.edu.sg.}, 
Ying-Chang Liang, \textit{Fellow, IEEE}\thanks{Y.-C.~Liang is with the Center for Intelligent Networking and Communications (CINC), University of Electronic Science and Technology of China (UESTC), Chengdu 611731, China. E-mail: liangyc@ieee.org.}, 
Qiang Yang, \textit{Fellow, IEEE}\thanks{Q. Yang is with Hong Kong University of Science and Technology, Hong Kong, China. Email: qyang@cse.ust.hk.}, 
Dusit Niyato, \textit{Fellow, IEEE},  
and Chunyan Miao
}

\maketitle

%====================================================================
\begin{abstract}

In recent years, mobile devices are equipped with increasingly advanced sensing and computing capabilities. Coupled with advancements in Deep Learning (DL), this opens up countless possibilities for meaningful applications, e.g., for medical purposes and in vehicular networks. Traditional cloud-based Machine Learning (ML) approaches require the data to be centralized in a cloud server or data center. However, this results in critical issues related to unacceptable latency and communication inefficiency. To this end, Mobile Edge Computing (MEC) has been proposed to bring intelligence closer to the edge, where data is produced. However, conventional enabling technologies for ML at mobile edge networks still require personal data to be shared with external parties, e.g., edge servers. Recently, in light of increasingly stringent data privacy legislations and growing privacy concerns, the concept of Federated Learning (FL) has been introduced. In FL, end devices use their local data to train an ML model required by the server. The end devices then send the model updates rather than raw data to the server for aggregation. FL can serve as an enabling technology in mobile edge networks since it enables the collaborative training of an ML model and also enables DL for mobile edge network optimization. However, in a large-scale and complex mobile edge network, heterogeneous devices with varying constraints are involved. This raises challenges of communication costs, resource allocation, and privacy and security in the implementation of FL at scale. In this survey, we begin with an introduction to the background and fundamentals of FL. Then, we highlight the aforementioned challenges of FL implementation and review existing solutions. Furthermore, we present the applications of FL for mobile edge network optimization. Finally, we discuss the important challenges and future research directions in FL.

\end{abstract}

%main

%====================================================================
%====================================================================
\section{Introduction}
\label{sec:intro}
%\cite{zhu2018cloud} \cite{wang2018deep} \cite{plastiras2018edge} \cite{lee2018techology} \cite{zhu2018towards} \cite{park2018wireless}

Currently, there are nearly $7$ billion connected Internet of Things (IoT) devices\cite{Lueth2018} and $3$ billion smartphones around the world. These devices are equipped with increasingly advanced sensors, computing, and communication capabilities. As such, they can potentially be deployed for various crowdsensing tasks, e.g., for medical purposes \cite{pryss2015mobile} and air quality monitoring \cite{ganti2011mobile}. Coupled with the rise of Deep Learning (DL) \cite{lecun2015deep}, the wealth of data collected by end devices opens up countless possibilities for meaningful research and applications. 

In the traditional cloud-centric approach, data collected by mobile devices is uploaded and processed centrally in a cloud-based server or data center. In particular, data collected by IoT devices and smartphones such as measurements \cite{oletic2015design}, photos \cite{jing2017crowdtracker}, videos \cite{hong2016optimizing}, and location information \cite{he2014developing} are aggregated at the data center \cite{li2017multi}. Thereafter, the data is used to provide insights or produce effective inference models. However, this approach is no longer sustainable for the following reasons. Firstly, data owners are increasingly privacy sensitive. Following privacy concerns among consumers in the age of big data, policy makers have responded with the implementation of data privacy legislations such as the European Commission's General Data Protection Regulation (GDPR) \cite{custers2019eu} and Consumer Privacy Bill of Rights in the US \cite{gaff2014privacy}. In particular, the consent (GDPR Article 6) and data minimalization principle (GDPR Article 5) limits data collection and storage only to what is consumer-consented and absolutely necessary for processing. Secondly, a cloud-centric approach involves long propagation delays and incurs unacceptable latency \cite{mao2017survey} for applications in which real-time decisions have to be made, e.g., in self-driving car systems \cite{ananthanarayanan2017real}. Thirdly, the transfer of data to the cloud for processing burdens the backbone networks especially in tasks involving unstructured data, e.g., in video analytics \cite{li2018learning}. This is exacerbated by the fact that cloud-centric training is relatively reliant on wireless communications \cite{han2019convergence}. As a result, this can potentially impede the development of new technologies. 

\begin{figure}[t!]
 \centering
\includegraphics[width=\columnwidth]{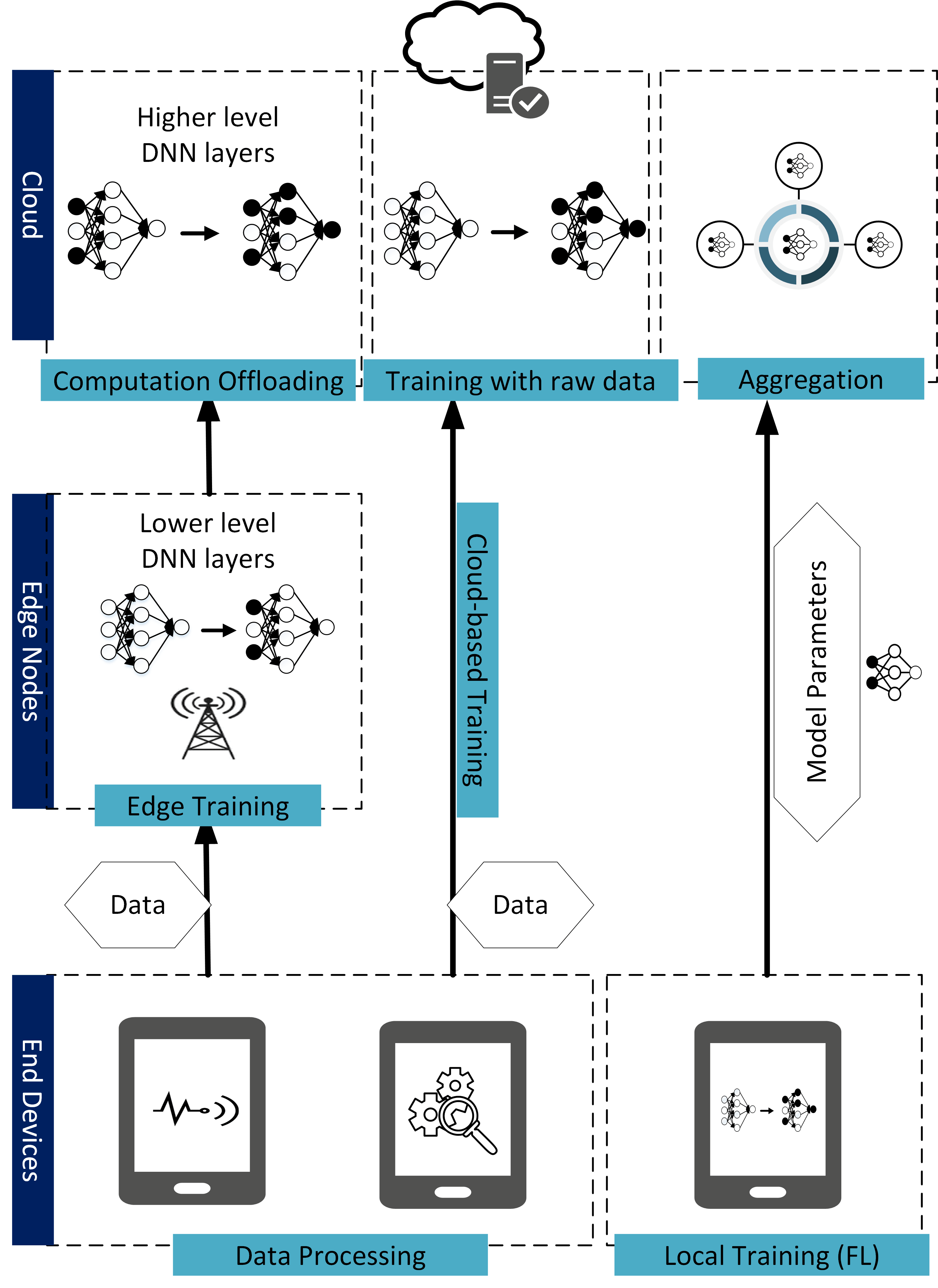}
 \caption{\small Edge AI approach brings AI processing closer to where data is produced. In particular, FL allows training on devices where the data is produced.}
 \label{Edge_AI}
\end{figure}

With data sources mainly located outside the cloud today \cite{cisco}, Mobile Edge Computing (MEC) has naturally been proposed as a solution in which the computing and storage capabilities \cite{mao2017survey} of end devices and edge servers are leveraged on to bring model training closer to where data is produced \cite{shi2016edge}. As defined in \cite{han2019convergence}, an end-edge-cloud computing network comprises: (i) end devices, (ii) edge nodes, and (iii) cloud server. For model training in conventional MEC approaches, a collaborative paradigm has been proposed in which training data are first sent to the edge servers for model training up to lower level DNN layers, before more computation intensive tasks are offloaded to the cloud \cite{chen2015efficient}, \cite{mach2017mobile} (Fig. \ref{Edge_AI}). However, this arrangement incurs significant communication costs and is unsuitable especially for applications that require persistent training \cite{han2019convergence}. In addition, computation offloading and data processing at edge servers still involve the transmission of potentially sensitive personal data. This can discourage privacy-sensitive consumers from taking part in model training, or even violate increasingly stringent privacy laws \cite{custers2019eu}. Although various privacy preservation methods, e.g.,
differential privacy (DP)~\cite{Abadi2016Deep}, have been proposed, a number of users are still not willing to expose their private
data for fear that their data may be inspected by external servers. In the long run, this discourages the
development of technologies as well as new applications. 

To guarantee that training data remains on personal devices and to facilitate collaborative machine learning of complex models among distributed devices, a decentralized ML approach called Federated Learning (FL) is introduced in \cite{mcmahan2016federated}. In FL, mobile devices use their local data to cooperatively train an ML model required by an FL server. They then send the model updates, i.e., the model's weights, to the FL server for aggregation. The steps are repeated in multiple rounds until a desirable accuracy is achieved. This implies that FL can be an enabling technology for ML model training at mobile edge networks. As compared to conventional cloud-centric ML model training approaches, the implementation of FL for model training at mobile edge networks features the following advantages. 

\begin{itemize}

\item \textit{Highly efficient use of network bandwidth:} Less information is required to be transmitted to the cloud. For example, instead of sending the raw data over for processing, participating devices only send the updated model parameters for aggregation. As a result, this significantly reduces costs of data communication and relieves the burden on backbone networks. 
\item \textit{Privacy:} Following the above point, the raw data of users need not be sent to the cloud. Under the assumption that FL participants and servers are non-malicious, this enhances user privacy and reduces the probability of eavesdropping to a certain extent. In fact, with enhanced privacy, more users will be willing to take part in collaborative model training and so, better inference models can be built.
\item \textit{Low latency:} With FL, ML models can be consistently trained and updated. Meanwhile, in the MEC paradigm, real-time decisions, e.g., event detection\cite{cicirelli2017edge}, can be made locally at the edge nodes or end devices. Therefore, the latency is much lower than that when decisions are made in the cloud before transmitting them to the end devices. This is vital for time critical applications such as self-driving car systems in which the slightest delays can potentially be life threatening \cite{ananthanarayanan2017real}. 

\end{itemize}

Given the aforementioned advantages, FL has seen recent successes in several applications. For example, the Federated Averaging algorithm (\textit{FedAvg}) proposed in \cite{mcmahan2016communication} has been applied to Google's Gboard \cite{hard2018federated} to improve next-word prediction models. In addition, several studies have also explored the use of FL in a number of scenarios in which data is sensitive in nature, e.g., to develop predictive models for diagnosis in health AI \cite{brisimi2018federated} and to foster collaboration across multiple hospitals \cite{nvidia} and Government agencies \cite{verma2018federated}.

Besides being an enabling technology for ML model training \textit{at} mobile edge networks, FL has also been increasingly applied as an enabling technology \textit{for} mobile edge network optimization. Given the computation and storage constraints of increasingly complex mobile edge networks, conventional network optimization approaches that are built on static models fare relatively poorly in modelling dynamic networks \cite{han2019convergence}. As such, a data-driven Deep Learning (DL) based approach \cite{simeone2018very} for optimizing resource allocation is increasingly popular. For example, DL can be used for representation learning of network conditions \cite{zhang2019deep} whereas Deep Reinforcement Learning (DRL) can optimize decision making through interactions with the dynamic environment \cite{luong2019applications}. However, the aforementioned approaches require user data as an input and these data may be sensitive or inaccessible in nature due to regulatory constraints. As such, in this survey, we also consider FL's potential to serve as an enabling technology for optimizing mobile edge networks, e.g., in cell association \cite{hamidouche2018collaborative}, computation offloading \cite{wang2018edge}, and vehicular networks \cite{samarakoon2018distributed}.

However, there are several challenges to be solved before FL can be implemented at scale. Firstly, even though raw data no longer needs to be sent to the cloud servers, communication costs remain an issue due to the high dimensonality of model updates and limited communication bandwith of participating mobile devices. In particular, state-of-the art DNN model training can involve the communication of millions of parameters for aggregation. Secondly, in a large and complex mobile edge network, the heterogeneity of participating devices in terms of data quality, computation power, and willingness to participate have to be well managed from the resource allocation perspective. Thirdly, FL does not guarantee privacy in the presence of malicious participants or aggregating servers. In particular, recent research works have clearly shown that a malicious participant may exist in FL and can infer the information of other participants just from the shared parameters alone. As such, privacy and security issues in FL still need to be considered.

Although there are surveys on MEC and FL, the existing studies usually treat the two topics separately. For existing surveys on FL, the authors in \cite{yang2019federated} place more emphasis on discussing the architecture and categorization of different FL settings to be used for the varying distributions of training data. The authors in \cite{niknam2019federated} highlight the applications of FL in wireless communications but do not discuss the issues pertaining to FL implementation. In addition, the focus of \cite{niknam2019federated} is on cellular network architecture rather than mobile edge networks. In contrast, the authors in \cite{li2019federated} provide a brief tutorial on FL and the challenges related to its implementation, but do not consider the issue of resource allocation in FL, or the potential applications of FL for mobile edge network optimization. On the other hand, for surveys in MEC that focus on implementing ML model training at edge networks, a macroscopic approach is usually adopted in which FL is briefly mentioned as one of the enabling technologies in the MEC paradigm, but without detailed elaboration with regards to its implementation or the related challenges. In particular, the authors in \cite{li2018learning}, \cite{zhou2019edge}, and \cite{cui2018survey} study the architectures and process of training and inference at edge networks without considering the challenges to FL implementation. In addition, surveys studying the implementation of DL for mobile edge network optimization mostly do not focus on FL as a potential solution to preserve data privacy. For example, the authors in \cite{mao2017survey,mach2017mobile,kumar2013survey,abbas2017mobile,wang2017survey,yao2019mobile} discuss strategies for optimizing caching and computation offloading for mobile edge networks, but do not consider the use of privacy preserving federated approaches in their studies. Similarly, \cite{luong2019applications} considers the use of DRL in communications and networking but do not include federated DRL approaches.

\begin{table*}
\centering
\arrayrulecolor{black}
\caption{\small An overview of selected surveys in FL and MEC}
\label{surveytable}
\begin{tabular}{!{\color{black}\vrule}l|l!{\color{black}\vrule}l!{\color{black}\vrule}} 
\hline
\rowcolor[rgb]{0.682,0.667,0.667} \multicolumn{1}{|l|}{\textbf{Ref.}} & \textbf{Subject}      & \textbf{Contribution}                                                                                                                \\ 
\hline
\cite{yang2019federated}                                                                    & \multirow{3}{*}{FL}   & Introductory tutorial on categorization of different FL settings, e.g., vertical FL, horizontal FL, and Federated Transfer Learning  \\ 
\cline{1-1}\cline{3-3}
\cite{niknam2019federated}                                                                   &                       & FL in optimizing resource allocation for wireless networks while preserving data privacy                                             \\ 
\cline{1-1}\cline{3-3}
\cite{li2019federated}                                                                    &                       & Tutorial on FL and discussions of implementation challenges in FL                                                                    \\ 
\cline{1-1}\arrayrulecolor{black}\cline{2-2}\arrayrulecolor{black}\cline{3-3}
\cite{li2018learning}                                                                   & \multirow{10}{*}{MEC} & Computation offloading strategy to optimize DL performance in edge computing                                                         \\ 
\cline{1-1}\cline{3-3}
\cite{zhou2019edge}                                                                   &                       & Survey on architectures and frameworks for edge intelligence                                                                         \\ 
\cline{1-1}\cline{3-3}
\cite{cui2018survey}                                                                  &                       & ML for IoT management, e.g., network management and security                                                                         \\ 
\cline{1-1}\cline{3-3}
\cite{mach2017mobile}                                                                   &                       & Survey on computation offloading in MEC                                                                                              \\ 
\cline{1-1}\cline{3-3}
\cite{luong2019applications}                                                                     &                       & Survey on DRL approaches to address issues in communications and networking                                                          \\ 
\cline{1-1}\cline{3-3}
\cite{kumar2013survey}                                                                    &                       & Survey on techniques for computation offloading                                                                                      \\ 
\cline{1-1}\cline{3-3}
\cite{abbas2017mobile}                                                                   &                       & Survey on architectures and applications of MEC                                                                                      \\ 
\cline{1-1}\cline{3-3}
\cite{wang2017survey}                                                                    &                       & Survey on computing, caching, and communication issues at mobile edge networks                                                       \\ 
\cline{1-1}\cline{3-3}
\cite{yao2019mobile}                                                                   &                       & Survey on the phases of caching and comparison among the different caching schemes                                                   \\ 
\cline{1-1}\cline{3-3}
\cite{mao2017survey}                                                                    &                       & Survey on joint mobile computing and wireless communication resource management in MEC                                               \\
\hline
\end{tabular}
\end{table*}

\begin{figure*}[!]
 \centering
\includegraphics[width=\linewidth]{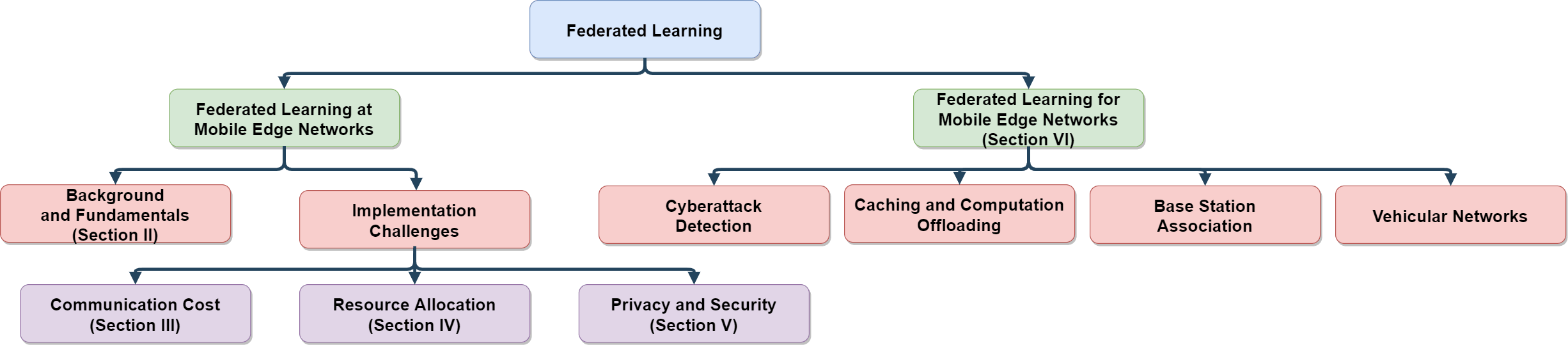}
 \caption{\small Classification of related studies to be discussed in this survey. }
 \label{flowchart}
\end{figure*}

In summary, most existing surveys on FL do not consider the applications of FL in the context of mobile edge networks, whereas existing surveys on MEC do not consider the challenges to FL implementation, or the potential of FL to be applied in mobile edge network optimization. This motivates us to have a comprehensive survey that has the following contributions: 

\begin{itemize}

\item We motivate the importance of FL as an important paradigm shift towards enabling collaborative ML model training. Then, we provide a concise tutorial on FL implementation and present to the reader a list of useful open-source frameworks that paves the way for future research on FL and its applications. 

\item We discuss the unique features of FL relative to a centralized ML approach and the resulting implementation challenges. For each of this challenge, we present to the reader a comprehensive discussion of existing solutions and approaches explored in the FL literature.

\item We discuss FL as an enabling technology for mobile edge network optimization. In particular, we discuss the current and potential applications of FL as a privacy-preserving approach for applications in edge computing.

\item We discuss the challenges and future research directions of FL.

\end{itemize}

For the reader's convenience, we classify the related studies to be discussed in this survey in Fig. \ref{flowchart}. The classification is based on (i) FL at mobile edge network, i.e., studies that focus on solving the challenges and issues related to implementing the collaborative training of ML models on end devices, and (ii) FL for mobile edge network, i.e., studies that specifically explore the application of FL for mobile edge network optimization. While the former group of studies works on addressing the fundamental issues of FL, the latter group uses FL as an application tool to solve issues in edge computing. We also present a list of common abbreviations for reference in Table \ref{tab:abbrev}.

The rest of this paper is organized as follows. Section \ref{sec:intro_FD} introduces the background and fundamentals of FL. Section \ref{sec: communication} reviews solutions provided to reduce communication costs. Section \ref{sec:resource} discusses resource allocation approaches in FL. Section \ref{sec:security} discusses privacy and security issues. Section \ref{sec:application} discusses applications of FL for mobile edge network optimization. Section \ref{sec:challenges_open_issues} discusses the challenges and future research directions in FL. Section \ref{sec:conclusions} concludes the paper.

%====================================================================
%====================================================================

\begin{table}[]
\centering{}\caption{\small List of common abbreviations. \label{tab:abbrev}}
\begin{tabular}{|l|l|}
\hline
\rowcolor[HTML]{9B9B9B} 
\textbf{Abbreviation} & \textbf{Description}                                                                    \\ \hline
BAA                   & Broadband Analog Aggregation                                                            \\ \hline
CNN                   & Convolutional Neural Network                                                            \\ \hline
CV                    & Computer Vision                                                                         \\ \hline
DDQN                  & Double Deep Q-Network                                                                   \\ \hline
DL                    & Deep Learning                                                                           \\ \hline
DNN                   & Deep Neural Network                                                                     \\ \hline
DP                    & Differential Privacy                                                                    \\ \hline
DQL                   & Deep Q-Learning                                                                         \\ \hline
DRL                   & Deep Reinforcement Learning                                                             \\ \hline
FedAvg                & Federated Averaging                                                                     \\ \hline
FL                    & Federated Learning                                                                      \\ \hline
GAN                   & Generative Adversarial Network                                                          \\ \hline
IID                   & \begin{tabular}[c]{@{}l@{}}Independent and Identically \\ Distributed\end{tabular}      \\ \hline
IoT                   & Internet of Things                                                                      \\ \hline
IoV                   & Internet of Vehicles                                                                    \\ \hline
LSTM                  & Long Short Term Memory                                                                  \\ \hline
MEC                   & Mobile Edge Computing                                                                   \\ \hline
ML                    & Machine Learning                                                                        \\ \hline
MLP                   & Multilayer Perceptron                                                                   \\ \hline
NLP                   & Natural Language Processing                                                             \\ \hline
OFDMA                 & \begin{tabular}[c]{@{}l@{}}Orthogonal Frequency-division\\ Multiple Access\end{tabular} \\ \hline
QoE                   & Quality of Experience                                                                   \\ \hline
RNN                   & Recurrent Neural Network                                                                \\ \hline
SGD                   & Stochastic Gradient Descent                                                             \\ \hline
SMPC                  & Secure Multiparty Computation                                                           \\ \hline
SNR                   & Signal-to-noise ratio                                                                   \\ \hline
SVM                   & Support Vector Machine                                                                  \\ \hline
TFF                   & TensorFlow Federated                                                                    \\ \hline
UE                    & User Equipment                                                                          \\ \hline
URLLC                 & Ultra reliable low latency communication                                                \\ \hline
\end{tabular}
\end{table}
\section{Background and Fundamentals of Federated Learning}
\label{sec:intro_FD}

\label{sec:intro_edge_AI}

Artificial Intelligence (AI) has become an essential part of our lives today, following the recent successes and progression of DL in several domains, e.g., Computer Vision (CV) \cite{dong2014learning} and Natural Language Processing (NLP) \cite{schmidhuber2015deep}. In traditional training of Deep Neural Networks (DNNs), a cloud based approach is adopted whereby data is centralized and model training occurs in powerful cloud servers. However, given the ubiquity of mobile devices that are equipped with increasingly advanced sensing and computing capabilities, the trend of migrating intelligence from the cloud to the edge, i.e., in the MEC paradigm, has naturally arisen. In addition, amid growing privacy concerns, the concept of FL has been proposed. 

FL involves the collaborative training of DNN models on end devices. There are, in general, two steps in the FL training process namely (i) local model training on end devices and (ii) global aggregation of updated parameters in the FL server. In this section, we first provide a brief introduction to DNN model training, which generalizes local model training in FL. Note that while FL can be applied to the training of ML models in general, we focus specifically on DNN model training in this section as a majority of the papers that we subsequently review study the federated training of DNN models. In addition, the DNN models are easily aggregated and outperform conventional ML techniques especially when the data is large. The implementation of FL at mobile edge networks can thus naturally leverage on the increasing computing power and wealth of data collected by distributed end devices, both of which are driving forces contributing to the rise of DL \cite{chen2014big}. As such, a brief introduction to general DNN model training will be useful for subsequent sections. Thereafter, we proceed to provide a tutorial of the FL training process that incorporates both global aggregation and local training. In addition, we also highlight the statistical challenges of FL model training and present the protocols and open-source frameworks of FL.

\subsection{Deep Learning}

Conventional ML algorithms rely on \textit{hand-engineered} feature extractors to process raw data \cite{trigeorgis2016adieu}. As such, domain expertise is often a prerequisite for building an effective ML model. In addition, feature selection has to be customized and reinitiated for each new problem. On the other hand, DNNs are representation learning based, i.e., DNNs can automatically discover and learn these features from raw data \cite{lecun2015deep} and thus often outperform conventional ML algorithms especially when there is an abundance of data. 

\begin{figure}[tbh]
\begin{centering}
\includegraphics[width=\columnwidth]{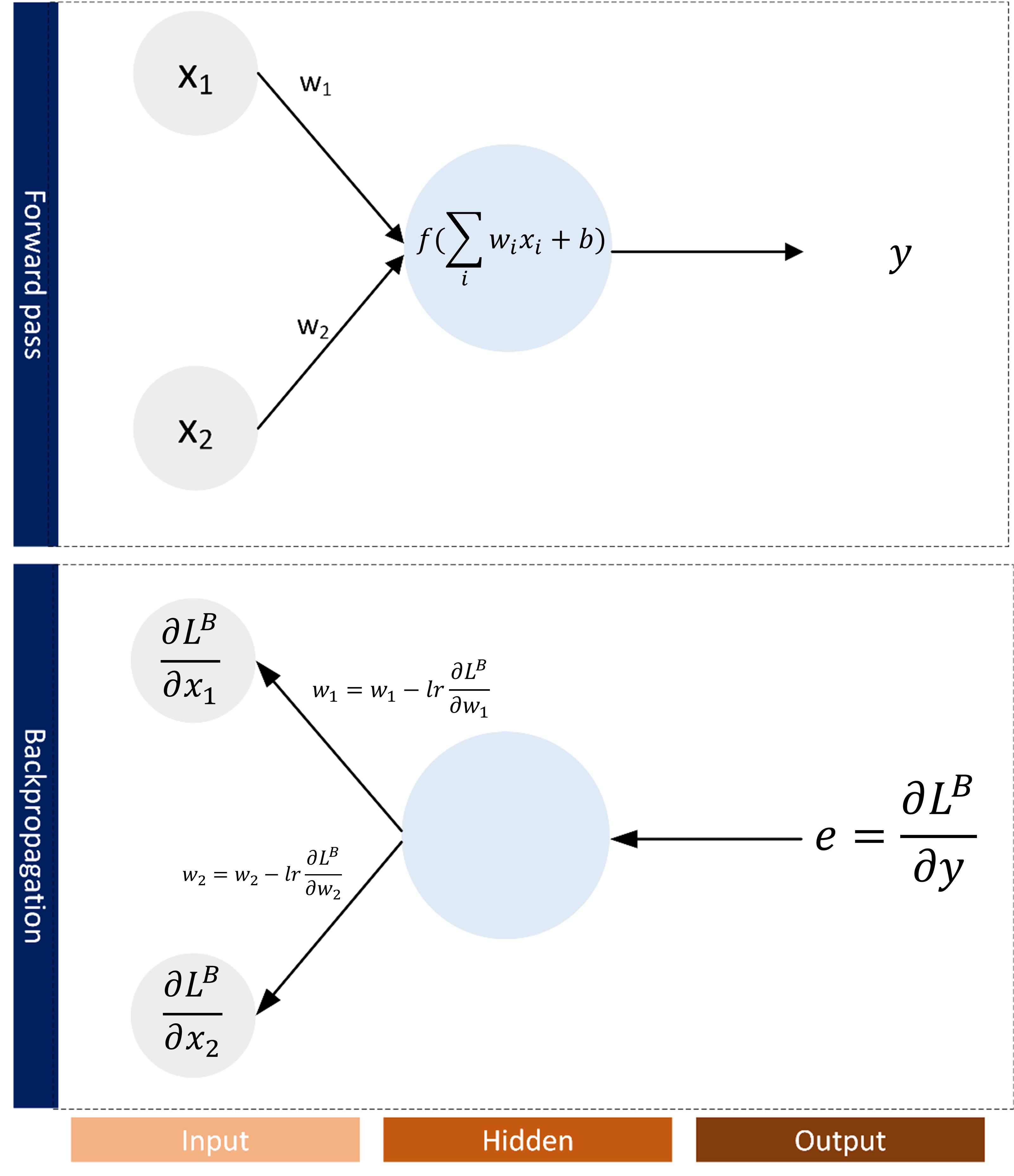}
\par\end{centering}
\caption{\small In forward pass, an output is derived from the weights and inputs. In backpropagation, the input gradient $e$ is used to calibrate the weights of the DNN model.\label{fig:backprop}}

\end{figure}

DL lies within the domain of the brain-inspired computing paradigm, of which the neural network is an important part of \cite{sze2017efficient}. In general, a neural network design emulates that of a neuron \cite{hecht1992theory}. It comprises three layers: (i) input layer, (ii) hidden layer, and (iii) output layer. In a conventional feedforward neural network, a weighted and bias-corrected input value is passed through a non-linear activation function to derive an output \cite{agostinelli2014learning} (Fig. \ref{fig:backprop}). Some activation functions include the ReLu and softmax functions \cite{schmidhuber2015deep}. A typical DNN comprises multiple hidden layers that map an input to an output. For example, the goal of a DNN trained for image classification \cite{xiao2017fashion} is to produce a vector of scores as the output, in which the positional index of the highest score corresponds to the class to which the input image is classified to belong. As such, the objective of training a DNN is to optimize the weights of the network such that the loss function, i.e., difference between the ground truth and model output, is minimized.

Before training, the dataset is first split into the training and test dataset. Then, the training dataset is used as input data for the optimization of weights in the DNN. The weights are calibrated through stochastic gradient descent (SGD), in which the weights are updated by the product of (i) the learning rate $lr$, i.e., the step size of gradient descent in each iteration, and (ii) partial derivative of the loss function $L$ with respect to the weight $w$. The SGD formula is as follows:
\begin{align}
W = W - lr \frac{\partial L}{\partial W} \label{eq:sgd_dl_1}\\
\frac{\partial L}{\partial W} \approx \frac{1}{m} \sum_{i\epsilon B} \frac{\partial l^{(i)}}{\partial W}\label{eq:sgd_dl_2}
\end{align}

Note that the SGD formula presented in (\ref{eq:sgd_dl_1}) is that of a mini-batch GD. In particular, equation (\ref{eq:sgd_dl_2}) is derived as the average gradient matrix over the gradient matrices of $B$ batches, in which each batch is a random subset consisting of $m$ training samples. This is preferred over the full batch GD, i.e., where the entirety of the training set is included in computing the partial derivative, since the full batch GD can lead to slow training and batch memorization \cite{hinton2012neural}. The gradient matrices are derived through backpropagation from the input gradient $e$ (Fig. \ref{fig:backprop}) \cite{hecht1992theory}.

The training iterations are then repeated over many epochs, i.e., full passes over the training set, for loss minimalization. A well-trained DNN generalizes well, i.e., achieve high \textit{inference} accuracy when applied to data that it has not seen before, e.g., the test set. There are other alternatives to supervised learning, e.g., semi-supervised learning \cite{zhu2005semi}, unsupervised learning \cite{radford2015unsupervised} and reinforcement learning \cite{mnih2013playing}. In addition, there also exists several DNN networks and architectures tailored to process the varying natures of input data, e.g., Multilayer Perceptron (MLP) \cite{bourlard1988auto}, Convolutional Neural Network (CNN) \cite{krizhevsky2012imagenet} typically for CV tasks, and Recurrent Neural Network (RNN) \cite{mikolov2010recurrent} usually for sequential tasks. However, an in-depth discussion is out of the scope of this paper. We refer interested readers to \cite{zhang2018survey, arulkumaran2017brief, han2018advanced, zhao2017survey, wang2019deep, arulkumaran2017deep} for comprehensive discussions of DNN architectures and training strategies. We next focus on FL, an important paradigm shift towards enabling privacy preserving and collaborative DL model training.

%====================================================================
%===================================================================

\subsection{Federated Learning}
\label{fltrain}

Motivated by privacy concerns among data owners, the concept of FL is introduced in \cite{mcmahan2016federated}. FL allows users to collaboratively train a
shared model while keeping personal data on their devices, thus alleviating
their privacy concerns. As such, FL can serve as an enabling technology for ML model training at mobile edge networks. For an introduction to the categorizations of different FL settings, e.g., vertical and horizontal FL, we refer the interested readers to \cite{yang2019federated}.

In general, there are two main entities in the FL
system, i.e., the data owners (viz. \textit{participants}) and the model owner (viz. \textit{FL server}). Let $\mathcal{N}=\left\{ 1,\ldots,N\right\} $
denote the set of $N$ data owners, each of which has
a private dataset $D_{i\in\mathcal{N}}$. Each data owner $i$ uses
its dataset $D_{i}$ to train a\emph{ local model} $\mathbf{w}_{i}$
and send only the local model parameters to the FL server. Then,
all collected local models are aggregated $\mathbf{\mathbf{w}}=\cup_{i\in\mathcal{N}}\mathbf{w}_{i}$
to generate a \emph{global model} $\mathbf{w}_{G}$. This
is different from the traditional centralized training which uses
$\mathbf{D}=\cup_{i\in\mathcal{N}}D_{i}$ to train a model $\mathbf{w}_{T}$, i.e., data from each individual source is aggregated first before model training takes place centrally.

\begin{figure}[tbh]
\begin{centering}
\includegraphics[width=\columnwidth]{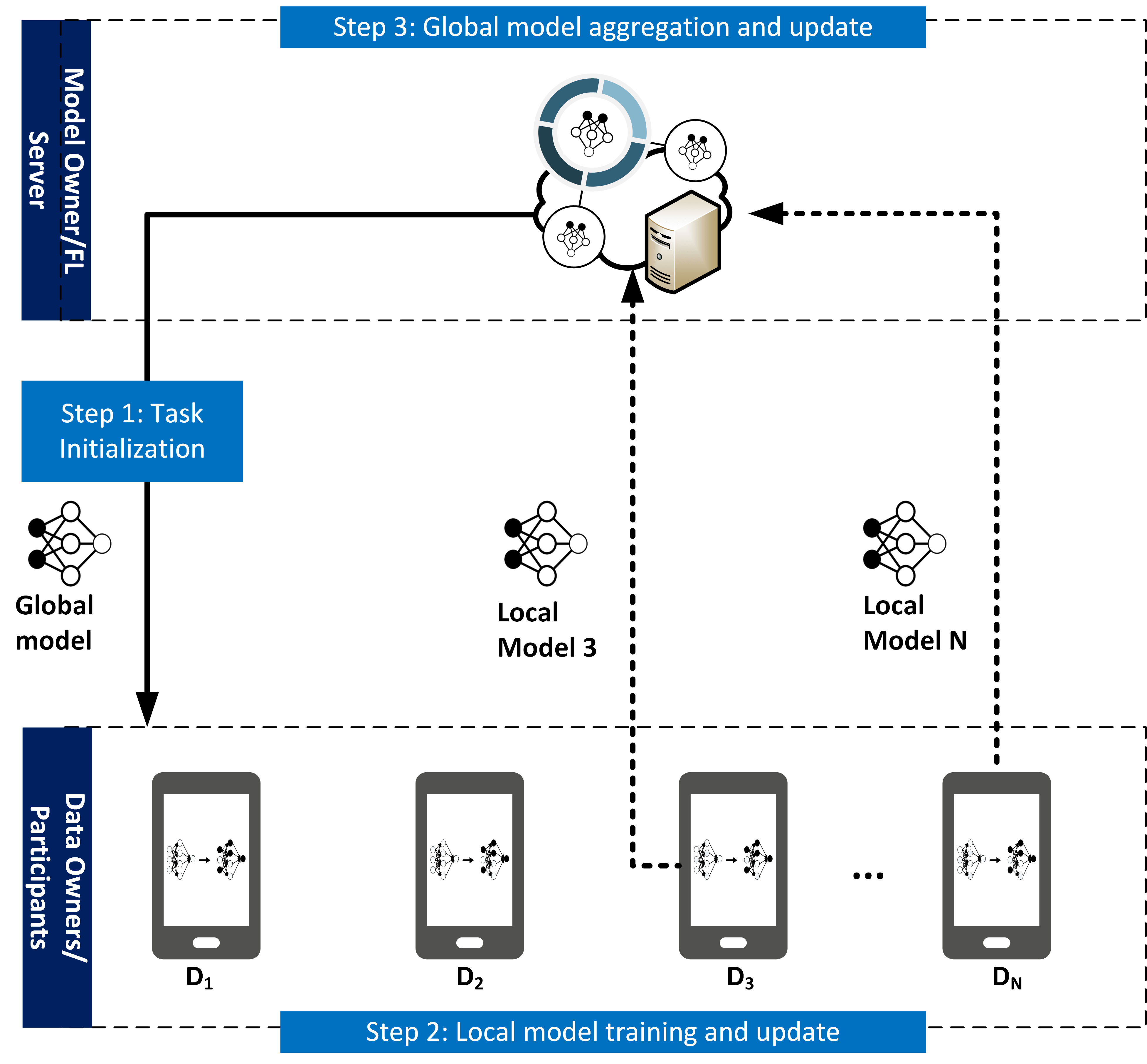}
\par\end{centering}
\caption{\small General FL training process involving \textit{N} participants.\label{fig:Federated-learning-model}}

\end{figure}
A typical architecture and training process of an FL system is shown in
Fig.~\ref{fig:Federated-learning-model}. In this system, the data
owners serve as the FL participants which collaboratively
train an ML model required by an aggregate server. An underlying assumption is that the data owners are honest, which
means they use their real private data to do the training and submit the
true local models to the FL server. Of course, this assumption may not always be realistic \cite{kangincentive2} and we discuss the proposed solutions subsequently in Sections \ref{sec:resource} and \ref{sec:security}. 

In general,
the FL training process includes the following three steps.
Note: the \textit{local} model refers to
the model trained at each participating device, whereas the \textit{global} model refers to
the model aggregated by the FL server. 
\begin{itemize}
\item \textit{Step 1 (Task initialization)}: The server decides the training task, i.e., the target application, and the corresponding data requirements.
The server also specifies the hyperparameters of the global model
and the training process, e.g., learning rate. Then, the server broadcasts the initialized
global model $\mathbf{w}_{G}^{0}$ and task to selected participants. 
\item \textit{Step 2 (Local model training and update)}: Based on the global model
$\mathbf{w}_{G}^{t}$, where $t$ denotes the current iteration index,
each participant respectively uses its local data and device to update
the local model parameters $\mathbf{w}_{i}^{t}$. The goal of participant $i$ in iteration \emph{$t$}
is to find optimal parameters $\mathbf{w}_{i}^{t}$ that minimize
the loss function $L(\mathbf{w}_{i}^{t})$, i.e., 
\begin{equation}
\mathbf{w}_{i}^{t^{*}}=\arg\min_{\mathbf{w}_{i}^{t}}L(\mathbf{w}_{i}^{t}).\label{eq:local_training_goal}
\end{equation}
The updated local model parameters are subsequently sent
to the server. 
\item \textit{Step 3 (Global model aggregation and update)}: The server aggregates
the local models from participants and then sends the updated
global model parameters $\mathbf{w}_{G}^{t+1}$ back to the data owners.
\end{itemize}

The server wants to minimize the global loss function $L(\mathbf{w}_{G}^{t})$, i.e., 
\begin{equation}
L(\mathbf{w}_{G}^{t})=\frac{1}{N}\sum_{i=1}^{N}L(\mathbf{w}_{i}^{t}).\label{eq:global_goal}
\end{equation}
Steps $2$-$3$ are repeated until the global loss function converges or a desirable training accuracy is achieved.

 Note
that the FL training process can be used for different ML
models that essentially use the SGD method such as Support
Vector Machines (SVMs) \cite{burges1998tutorial}, neural networks, and linear regression \cite{myers1990classical}.
A training dataset usually contains a set of $n$ data feature vectors
$\mathbf{x}=\{\mathbf{x}_{1},\ldots,\mathbf{x}_{n}\}$ and a set of
corresponding data labels\footnote{In the case of unsupervised learning, there is no data label.}
$\mathbf{y}=\{y_{1},\ldots,y_{n}\}$. In addition, let $\hat{y_{j}}=f(\mathbf{x}_{j};\mathbf{w})$
denote the predicted result from the model $\mathbf{w}$ updated/trained by data
vector $x_{j}$. Table~\ref{tab:loss-functions} summarizes several
loss functions of common ML models~\cite{wang2019adaptive}. 

\begin{table}[tbh]
\centering{}\caption{\small Loss functions of common ML models\label{tab:loss-functions}}
\begin{tabular}{|>{\centering}m{1.9cm}|>{\centering}m{5cm}|}
\hline 
Model & Loss function $L(\mathbf{w}_{i}^{t})$\tabularnewline
\hline 
Neural network & $\frac{1}{n}\sum_{j=1}^{n}(y_{i}-f(\mathbf{x}_{j};\mathbf{w}))^{2}$\linebreak (Mean
Squared Error)\tabularnewline
\hline 
Linear regression & $\frac{1}{2}\left\Vert y_{j}-\mathbf{w}^{T}\mathbf{x}_{j}\right\Vert ^{2}$\tabularnewline
\hline 
K-means & $\sum_{j}\left\Vert \mathbf{x}_{j}-f(\mathbf{x}_{j};\mathbf{w})\right\Vert $\linebreak ($f(\mathbf{x}_{j};\mathbf{w})$
is the centroid of all objects assigned to $x_{j}$'s class)\tabularnewline
\hline 
squared-SVM & $[\frac{1}{n}\sum_{j=1}^{n}\max(0,1-y_{j}(\mathbf{w}^{T}\mathbf{x}_{j}-bias))]$\linebreak $+\lambda\left\Vert \mathbf{w}^{T}\right\Vert ^{2}$($bias$
is the bias parameter and $\lambda$ is const.)\tabularnewline
\hline 
\end{tabular}
\end{table}

Global model aggregation is an integral part of FL. A straightforward
and classical algorithm for aggregating the local models is the \textit{FedAvg} algorithm proposed in \cite{mcmahan2016communication}, which is similar to that of local SGD \cite{lin2018don}. The pseudocode for \textit{FedAvg} is given in Algorithm~\ref{alg:AveragingAlg}.
\begin{algorithm}[tbh]
\scriptsize
\begin{algorithmic}[1]
\Require{Local minibatch size $B$, number of participants $m$ per iteration, number of local epochs $E$, and learning rate $\eta$.}
\Ensure{Global model $\mathbf{w}_{G}$.}
\State{[Participant $i$]}
\State{\textbf{LocalTraining}($i$, $\mathbf{w}$):}
	\State{Split local dataset $D_i$ to minibatches of size $B$ which are included into the set $\mathcal{B}_i$.}
	\For{each local epoch $j$ from $1$ to $E$}
		\For{each $b \in \mathcal{B}_i$}
			\State{$\mathbf{w} \gets \mathbf{w} - \eta\Delta L(\mathbf{w};b)$ \qquad ($\eta$ is the learning rate and $\Delta L$ is the gradient of $L$ on $b$.)}
		\EndFor
	\EndFor
	\State{}
	\State{[Server]}
	\State{Initialize $\mathbf{w}_{G}^{0}$}
	\For{each iteration $t$ from $1$ to $T$}
		\State{Randomly choose a subset $\mathcal{S}_t$ of $m$ participants from $\mathcal{N}$}
		\For{each partipant $i \in \mathcal{S}_t$ $\textbf{parallely}$}
			\State{$\mathbf{w}_{i}^{t+1} \gets \textbf{LocalTraining}$($i$, $\mathbf{w}_{G}^{t}$)}
		\EndFor
\State{$\mathbf{w}_{G}^{t}=\frac{1}{\sum_{i\in\mathcal{N}}D_{i}}\sum_{i=1}^{N}D_{i}\mathbf{w}_{i}^{t}$ \qquad (Averaging aggregation)} 
	\EndFor

\end{algorithmic} 

\caption{Federated averaging algorithm~\cite{mcmahan2016communication}\label{alg:AveragingAlg}}
\end{algorithm}
 As described in Step 1 above, the server first initializes the task (lines 11-16). Thereafter, in Step 2, the participant $i$ implements the local
training and optimizes the target in (\ref{eq:local_training_goal})
on minibatches from the original local dataset (lines 2-8). Note that a minibatch refers to a randomized subset of each participant's dataset. At the $t^{th}$
iteration (line 17), the server minimizes the global loss in (\ref{eq:global_goal})
by the averaging aggregation which is formally defined as
\begin{equation}
\mathbf{w}_{G}^{t}=\frac{1}{\sum_{i\in\mathcal{N}}D_{i}}\sum_{i=1}^{N}D_{i}\mathbf{w}_{i}^{t}.\label{eq:averaging_aggr}
\end{equation}
The FL training process is iterated till the global loss function converges, or a desirable accuracy is achieved. 

\subsection{Statistical Challenges of FL}
\label{stats}

Following an elaboration of the FL training process in the previous section, we now proceed to discuss the statistical challenges faced in FL. 

In traditional distributed ML, the central server has access to
the whole training dataset. As such, the server can split the dataset
into subsets that follow similar distributions. The subsets are subsequently sent to
participating nodes for distributed training. However, this approach is
impractical for FL since the local dataset is only accessible by the data
owner. 

In the FL setting, the participants may have local datasets that follow different distributions, i.e., the datasets of participants are non-IID. While the authors in \cite{mcmahan2016communication} show that the aforementioned \textit{FedAvg} algorithm is able to achieve desirable accuracy even when data is non-IID across participants, the authors in \cite{zhao2018federated} found otherwise. For example, the accuracy of a \textit{FedAvg}-trained CNN model has 51\% lower accuracy than centrally-trained CNN model for CIFAR-10 \cite{krizhevsky2009learning}. This deterioration in accuracy is further shown to be quantified by the earth mover's distance (EMD) \cite{rubner2000earth}, i.e., difference in FL participant's data distribution as compared to the population distribution. As such, when data is non-IID and highly skewed, data-sharing is proposed in which a shared dataset with uniform distribution across all classes is sent by the FL server to each FL participant. Then, the participant trains its local model on its private data together with the received data. The simulation result shows that accuracy can be increased by 30\% with 5\% shared data due to reduced EMD.   However, a common dataset may not always be available for sharing by the FL server. An alternative solution to gather contributions towards building the common dataset is subsequently discussed in section \ref{sec:resource}.

The authors in \cite{duan2019astraea} also find that global imbalance, i.e., the situation in which the collection of data held across all FL participants is class imbalanced, leads to a deterioration in model accuracy. As such, the Astraea framework is proposed. On initialization, the FL participants first send their data distribution to the FL server. A rebalancing step is introduced before training begins in which each participant performs data augmentation \cite{wong2016understanding} on the minority classes, e.g., through random rotations and shifts. After training on the augmented data, a mediator is created to coordinate intermediate aggregation, i.e., before sending the updated parameters to the FL server for global aggregation. The mediator selects participants with data distributions that best contributes to an uniform distribution when aggregated. This is done through a greedy algorithm approach to minimize the Kullback-Leibler Divergence \cite{joyce2011kullback} between local data and uniform distribution. The simulation results show accuracy improvement when tested on imbalanced datasets.

Given the heterogeneity of data distribution across devices, there has been an increasing number of studies that borrow concepts from multi-task learning \cite{jaggi2014communication} to learn separate, but structurally related models for each participant. Instead of minimizing the conventional loss function presented previously in Table \ref{tab:loss-functions}, the loss function is modified to also model the relationship amongst tasks \cite{smith2017federated}. Then, the MOCHA algorithm is proposed in which an alternating optimization approach \cite{bezdek2003convergence} is used to approximately solve the minimization problem. Interestingly, MOCHA can also be calibrated based on the resource constraints of a participating device. For example, the quality of approximation can be adaptively adjusted based on network conditions and CPU states of the participating devices. However, MOCHA cannot be applied to non-convex DL models. 

Similarly, the authors in \cite{arivazhagan2019federated} also borrow concepts from multi-task learning to deal with the statistical heterogeneity in FL. The FEDPER approach is proposed in which all FL participants share a set of base layers trained using the \textit{FedAvg} algorithm. Then, each participant separately trains another set of personalization layers using its own local data. In particular, this approach is suitable for building recommender's systems given the diverse preferences of participants. The authors show empirically using the Flickr-AES dataset \cite{ren2017personalized} that the FEDPER approach outperforms a pure \textit{FedAvg} approach since the personalization layer is able to represent the personal preference of an FL participant. However, it is worth to note that the collaborative training of the base layers are still important to achieve a high test accuracy, since each participant has insufficient local data samples for purely personalized model training.

Apart from data heterogeneity, the convergence of a distributed learning algorithm is always
a concern. Higher convergence rate helps to save a large amount
of time and resources for the FL participants, and also significantly increases
the success rate of the federated training since fewer communication
rounds imply reduced participant dropouts. To ensure convergence, the study in \cite{Li2019} propose \textit{FedProx}, which modifies the loss function to also include a tunable parameter that restricts how much local updates can affect the prevailing model parameters. The \textit{FedProx} algorithm can be adaptively tuned, e.g., when training loss is increasing, model updates can be tuned to affect the current parameters less. Similarly, the authors in \cite{huang2018loadaboost} also propose the \textit{LoAdaBoost FedAvg} algorithm to complement the aforementioned data-sharing approach \cite{zhao2018federated} in ML on medical data. In \textit{LoAdaBoost FedAvg}, participants train the model on their local data and compare the cross-entropy loss with the median loss from the \textit{previous} training round. If the current cross-entropy loss is higher, the model is retrained before global aggregation so as to increase learning efficiency. The simulation results show that faster convergence is achieved as a result. 

In fact, the statistical challenges of FL coexist with other issues that we explore in subsequent sections. For example, the communication costs incurred in FL can be reduced by faster convergence. Similarly, resource allocation policies can also be designed to solve statistical heterogeneity. As such, we revisit these concepts in greater detail subsequently.

\subsection{FL protocols and frameworks}
\label{proto}

To improve scalability, an FL protocol 
has been proposed in \cite{bonawitz2019towards} from the system level. This protocol deals with
issues regarding unstable device connectivity and communication security etc.

\begin{figure*}[t]
\begin{centering}
\includegraphics[width=0.85\textwidth]{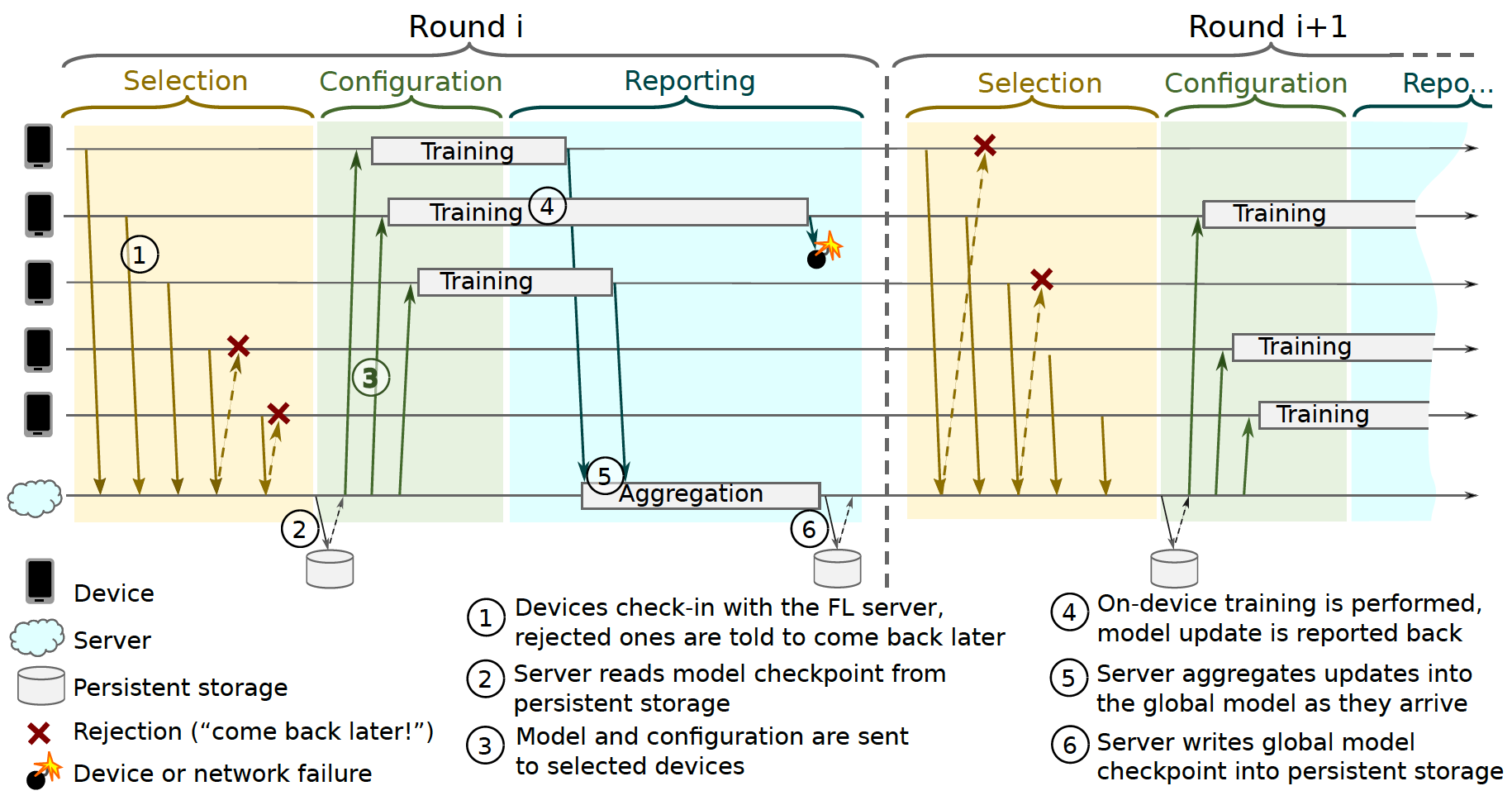}
\par\end{centering}
\caption{\small Federated learning protocol \cite{bonawitz2019towards}.\label{fig:Federated-learning-protocol}}
\end{figure*}

The FL protocol (Fig.~\ref{fig:Federated-learning-protocol}) consists of three phases in each training round: 

\begin{enumerate}

\item \emph{Selection:} In the participant selection phase, the FL server chooses a subset of connected devices to participate in a training round. The selection criteria may subsequently be calibrated to the server's needs, e.g., training efficiency \cite{nishio2018client}. In Section \ref{sec:resource}, we further elaborate on proposed participant selection methods.

\item \emph{Configuration:} The server is configured accordingly to the aggregation mechanism preferred, e.g. simple or secure aggregation \cite{bonawitz2017practical}. Then, the server sends the training schedule and global model to each participant.

\item \emph{Reporting:} The server receives updates from participants. Thereafter, the updates can be aggregated, e.g., using the \textit{FedAvg} algorithm.

\end{enumerate}

In addition, to manage device connections accordingly to varying FL population size, pace steering is also recommended. Pace steering adaptively manages the optimal time window for participants to reconnect to the FL server \cite{bonawitz2019towards}. For example, when the FL population is small, pace steering is used to ensure that there is a sufficient number of participating devices that connect to the server simultaneously. In contrast, when there is a large population, pace steering randomly chooses devices to participate to prevent the situation in which too many participating devices are connected at one point of time.

Apart from communication efficiency, communication security during local updates transmission is
another problem to be resolved. Specifically,
there are mainly two aspects in communication security:

\begin{enumerate}
\item \emph{Secure aggregation}: To prevent local updates from being traced and utilized
to infer the identity of the FL participant, 
a virtual and trusted third party server is deployed for local model
aggregation \cite{bonawitz2017practical}. The Secret Sharing mechanism~\cite{Bloemer2011} is also used for
transmission of local updates with authenticated encryption. 

\item \emph{Differential privacy}: Similar to secure aggregation, differential
privacy (DP) prevents the FL server from identifying the owner of a local
update. The difference is that to achieve the goal of privacy preservation,
the DP in FL~\cite{Mcmahan2018} adds a certain degree of noise in
the original local update while providing theoretical guarantees on
the model quality.
\end{enumerate}
These concepts on privacy and security are presented in detail in Section \ref{sec:security}. Recently, some open-source frameworks for FL have been developed as follows:

\begin{enumerate}

\item \emph{TensorFlow Federated (TFF)}: TFF \cite{tensorflow} is a framework based on Tensorflow developed by Google for decentralized ML and other distributed
computations. TFF consists of two layers (i) FL and (ii) Federated Core (FC). The FL layer is a high-level interface that allows the implementation of FL to existing TF models without the user having to apply the FL algorithms personally. The FC layer combines TF with communication operators to allow users to experiment with customized and newly designed FL algorithms.

\item \emph{PySyft}: PySyft~\cite{Ryffel2018} is a framework based on PyTorch
for performing encrypted, privacy-preserving DL and implementations of
related techniques, such as Secure Multiparty Computation (SMPC) and
DP, in untrusted environments while protecting data. Pysyft is developed such that it retains the native Torch interface, i.e., the ways to execute all tensor operations remain unchanged from that of Pytorch. When a SyftTensor is created, a LocalTensor is automatically created to also apply the input command to the native Pytorch tensor. To simulate FL, participants are created as \textit{Virtual Workers}. Data, i.e., in the structure of tensors, can be split and distributed to the Virtual Workers as a simulation of a practical FL setting. Then, a PointerTensor is created to specify the data owner and storage location. In addition, model updates can be fetched from the Virtual Workers for global aggregation.

\item \emph{LEAF}: An open source framework \cite{caldas2018leaf} of datasets that can be used as benchmarks in FL, e.g., Federated Extended MNIST (FEMNIST), an MNIST \cite{lecun2010mnist} dataset partitioned based on writer of each character, and Sentiment140 \cite{go2016sentiment140}, a dataset partitioned based on different users. In these datasets, the writer or user is assumed to be a participant in FL, and their corresponding data is taken to be the local data held in their personal devices. The implementation of newly designed algorithms on these benchmark datasets allow for reliable comparison across studies.

\item \emph{FATE}: Federated AI Technology Enabler (FATE) is an open-source framework by WeBank \cite{webank} that supports the federated and secure implementation of several ML models. 
\end{enumerate}

\subsection{Unique characteristics and issues of FL}

Besides the statistical challenges we present in Section \ref{stats}, FL has some unique characteristics and features \cite{Konecny2016} as compared to other distributed ML approaches:
\begin{enumerate}

\item \textit{Slow and unstable communication}: In the traditional distributed training
in a data center, the communication environment can be assumed to
be perfect where the information transmission rate is very high and
there is no packet loss. However, these assumptions are not applicable to 
the FL environment where heterogeneous devices are involved in training. For example, the Internet upload speed is typically
much slower than download speed \cite{konevcny2016federated}. Also, some participants with unstable wireless communication channels may consequently
drop out due to disconnection from the Internet.

\item \textit{Heterogeneous devices}: Apart from bandwidth constraints, FL involves heterogeneous devices with varying resource constraints. For example, the devices can have different computing capabilities, i.e., CPU states and battery level. The devices can also have different levels of \textit{willingness} to participate, i.e., FL training is resource consuming and given the distributed nature of training across numerous devices, there is a possibility of free ridership.

\item \textit{Privacy and security concerns}:
As we have previously discussed, data owners are increasingly privacy sensitive. However, as will be subsequently presented in Section \ref{sec:security}, malicious participants are able to infer sensitive information from shared parameters, which potentially negates privacy preservation. In addition, we have previously assumed that all participants and FL servers are trustful. In reality, they may be malicious.

\end{enumerate}
These unique characteristics of FL lead to several practical issues in FL implementation
mainly in three aspects that we now proceed to discuss, i.e., (i) statistical challenges (ii) communication costs (iii) resource allocation and (iv) privacy and security. In the following sections, we review related work that address each of these issues.

\section{Communication Cost}
\label{sec: communication}

In FL, a number of rounds of communications between the participants and the FL server may be required to achieve a target accuracy (Fig. \ref{fig:Federated-learning-protocol}). For complex DL model training involving, e.g. CNN, each update may comprise millions of parameters \cite{he2016deep}. The high dimensionality of the updates can result in the incurrence of high communication costs and can lead to a training bottleneck. In addition, the bottleneck can be worsened due to (i) unreliable network conditions of participating devices \cite{wangcmfl} and (ii) the asymmetry in Internet connection speeds in which upload speed is faster than download speed, resulting in delays in model uploads by participants \cite{konevcny2016federated}. As such, there is a need to improve the communication efficiency of FL. The following approaches to reduce communication costs are considered:

\begin{itemize}

\item \textit{Edge and End Computation}: In the FL setup, the communication cost often dominates computation cost \cite{mcmahan2016communication}. The reason is that on-device dataset is relatively small whereas mobile devices of participants have increasingly fast processors. On the other hand, participants may only be willing to participate in the model training only if they are connected to Wi-Fi \cite{konevcny2016federated}. As such, more computation can be performed on edge nodes or on end devices before each global aggregation so as to reduce the number of communication rounds needed for the model training. In addition, algorithms to ensure faster convergence can also reduce number of communication rounds involved, at the expense of more computation on edge servers and end devices.

\item \textit{Model Compression}: This is a technique commonly used in distributed learning \cite{wang2018atomo}. Model or gradient compression involves the communication of an update that is transformed to be more compact, e.g., through sparsification, quantization or subsampling \cite{stich2018sparsified}, rather than the communication of full update. However, since the compression may introduce noise, the objective is to reduce the size of update transferred during each round of communication while maintaining the quality of trained models \cite{caldas2018expanding}.
 
\item \textit{Importance-based Updating}: This strategy involves selective communication such that only the important or relevant updates \cite{tao2018esgd} are transmitted in each communication round. In fact, besides saving on communication costs, omitting some updates from participants can even improve the global model performance.

\end{itemize}

% \usepackage{booktabs}
% \usepackage{multirow}
% \usepackage{booktabs}
% \usepackage{multirow}
% \usepackage{multirow}
\subsection{Edge and End Computation}

\begin{figure}[!]
	\centering
	\includegraphics[width=\columnwidth]{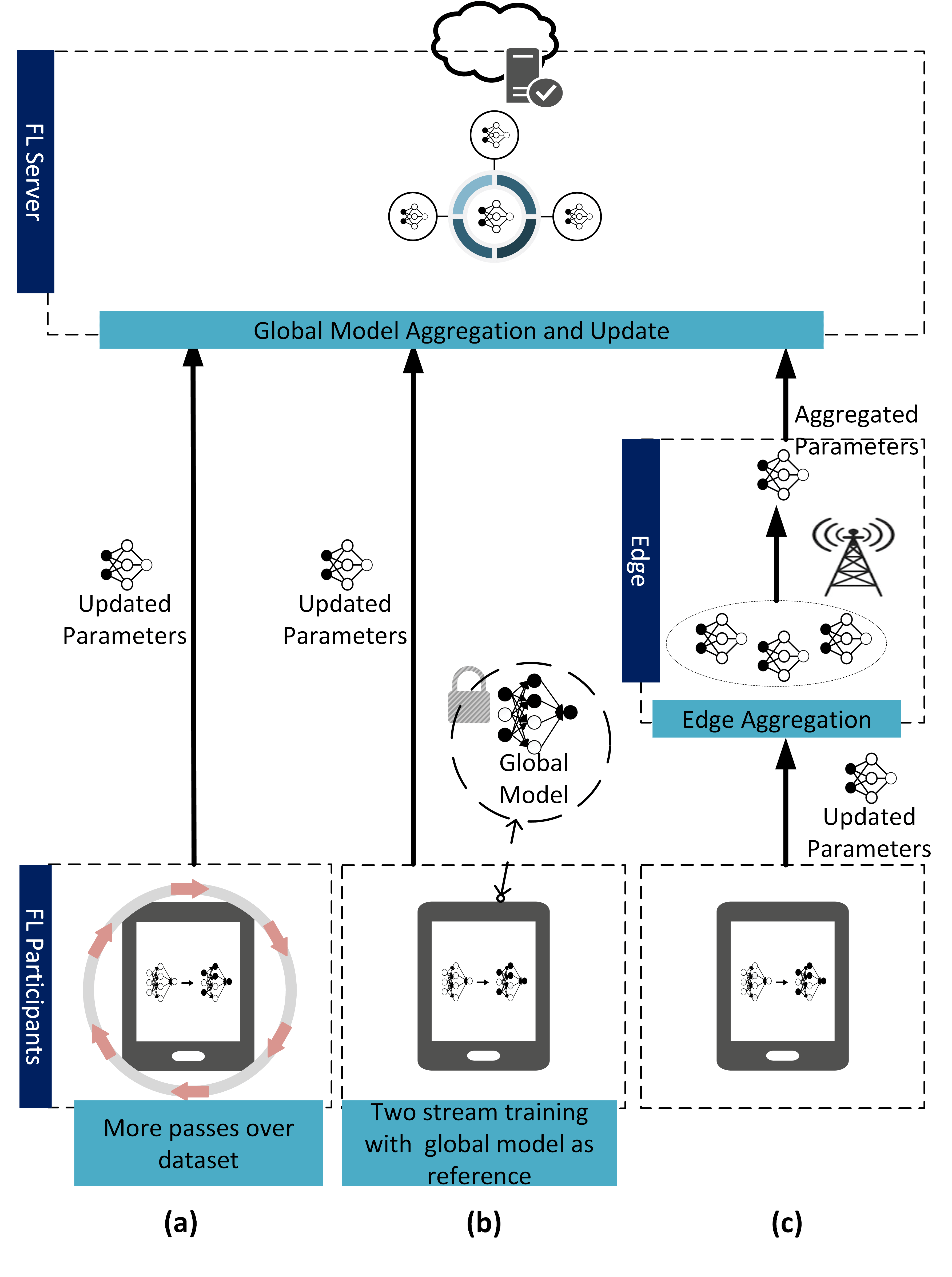}
	\caption{Approaches to increase computation at edge and end devices include (a) Increased computation at end devices, e.g., more passes over dataset before communication \cite{mcmahan2016communication,liu2019communication} (b) Two-stream training with global model as a reference \cite{yao2018two} and (c) Intermediate edge server aggregation \cite{liu2019edge}.}
	\label{fig:computation}
\end{figure}

To decrease the number of communication rounds, additional computation can be performed on participating end devices before each iteration of communication for global aggregation (Fig. \ref{fig:computation}(a)). The authors in \cite{mcmahan2016communication} consider two ways to increase computation on participating devices: (i) increasing parallelism in which more participants are selected to participate in each round of training and (ii) increasing computation per participant whereby each participant performs more local updates before communication for global aggregation. A comparison is conducted for the FederatedSGD (\textit{FedSGD}) algorithm and the proposed \textit{FedAvg} algorithm. For the \textit{FedSGD} algorithm, all participants are involved and only one pass is made per training round in which the minibatch size comprises of the entirety of the participant's dataset. This is similar to the full-batch training in centralized DL frameworks. For the proposed \textit{FedAvg} algorithm, the hyperparameters are tuned such that more local computations are performed by the participants. For example, the participant can make more passes over its dataset or use a smaller local minibatch size to increase computation before each communication round. The simulation results show that increased parallelism does not lead to significant improvements in reducing communication cost, once a certain threshold is reached. As such, more emphasis is placed on increasing computation per participant while keeping the fraction of selected participants constant. For MNIST CNN simulations, increased computation using the proposed \textit{FedAvg} algorithm can reduce communication rounds by more than $30$ times when the dataset is IID. For non-IID dataset, the improvement is less significant ($2.8$ times) using the same hyperparameters. However, for Long Short Term Memory (LSTM) simulations \cite{hochreiter1997long}, improvements are more significant even for non-IID data ($95.3$ times). In addition, \textit{FedAvg} increases the accuracy eventually since model averaging produces regularization effects similar to dropout \cite{srivastava2014dropout}, which prevents overfitting. 

As an extension, the authors in \cite{liu2019communication} also validate that a similar concept as that of \cite{mcmahan2016communication} works for vertical FL. In vertical FL, collaborative model training is conducted across the same set of participants with different data features. The Federated Stochastic Block Coordinate Descent (FedBCD) algorithm is proposed in which each participating device performs multiple local updates first before communication for global aggregation. In addition, convergence guarantee is also provided with an approximate calibration of the number of local updates per interval of communication.

Another way to decrease communication cost can also be through modifying the training algorithm to increase convergence speed, e.g., through the aforementioned \textit{LoAdaBoost FedAvg} in \cite{huang2018loadaboost}. Similarly, the authors in \cite{yao2018two} also propose increased computation on each participating device by adopting a two-stream model (Fig. \ref{fig:computation}(b)) commonly used in transfer learning and domain adaption \cite{long2015learning}. During each training round, the global model is received by the participant and fixed as a reference in the training process. During training, the participant learns not just from local data, but also from other participants with reference to the fixed global model. This is done through the incorporation of Maximum Mean Discrepancy (MMD) into the loss function. MMD measures the distance between the means of two data distributions \cite{long2015learning}, \cite{long2013transfer}. Through minimizing MMD loss between the local and global models, the participant can extract more generalized features from the global model, thus accelerating the convergence of training process and reducing communication rounds. The simulation results on the CIFAR-10 and MNIST dataset using DL models such as AlexNet\cite{krizhevsky2012imagenet} and 2-CNN respectively show that the proposed two-stream FL can reach the desirable test accuracy in 20\% fewer communication rounds even when data is non-IID. However, while convergence speed is increased, more computation resources have to be consumed by end devices for the aforementioned approaches. Given the energy constraints of participating mobile devices in particular, this necessitates resource allocation optimization that we subsequently discuss in Section \ref{sec:resource}.

While the aforementioned studies consider increasing computation on participating \textit{devices}, the authors in \cite{liu2019edge} propose an edge computing inspired paradigm in which proximate edge \textit{servers} can serve as intermediary parameter aggregators given that the propagation latency from participant to the edge server is smaller than that of the participant-cloud communication (Fig. \ref{fig:computation}(c)). A hierarchical FL (\textit{HierFAVG}) algorithm is proposed whereby for every few local participant updates, the edge server aggregates the collected local models. After a predefined number of edge server aggregations, the edge server communicates with the cloud for global model aggregation. As such, the communication between the participants and the cloud occurs only once after an interval of multiple local updates. Comparatively, for the \textit{FedAvg} algorithm proposed in \cite{mcmahan2016communication}, the global aggregation occurs more frequently since no intermediate edge server aggregation is involved. The authors further prove the covergence of \textit{HierFAVG} for both convex and non-convex objective functions given non-IID user data. The simulation results show that for the same number of local updates between two global aggregations, more intermediate edge aggregations before each global aggregation can lead to reduced communication overhead as compared to the \textit{FedAvg} algorithm. This result holds for both IID and non-IID data, implying that intermediate aggregation on edge servers may be implemented on top of \textit{FedAvg} so as to reduce communication costs. However, when applied to non-IID data, the simulation results show that \textit{HierFAVG} fails to converge to the desired accuracy level (90\%) in some instances, e.g., when edge-cloud divergence is large or when there are many edge servers involved. As such, a further study is required to better understand the tradeoffs between adjusting local and edge aggregation intervals, so as to ensure that the parameters of the \textit{HierFAVG} algorithm can be optimally calibrated to suit other settings. Nevertheless, \textit{HierFAVG} is a promising approach for the implementation of FL at mobile edge networks, since it leverages on the proximity of intermediate edge server to reduce communication costs, and potentially relieve the burden on the remote cloud.

\subsection{Model Compression}

To reduce communication costs, the authors in \cite{konevcny2016federated} propose structured and sketched updates to reduce the size of model updates sent from participants to the FL server during each communication round. 

\textit{Structured updates} restrict participant updates to have a pre-specified structure, i.e., low rank and random mask. For the low rank structure, each update is enforced to be a low rank matrix expressed as a product of two matrices. Here, one matrix is generated randomly and held constant during each communication round whereas the other is optimized. As such, only the optimized matrix needs to be sent to the server. For the random mask structure, each participant update is restricted to be a sparse matrix following a pre-defined random sparsity pattern generated independently during each round. As such, only the non-zero entries are required to be sent to the server. 

On the other hand, \textit{sketched updates} refer to the approach of encoding the update in a compressed form before communication with the server, which subsequently decodes the updates before aggregation. One example of sketched update is the subsampling approach, in which each participant communicates only a random subset of the update matrix. The server then averages the subsampled updates to derive an unbiased estimate of the true average. Another example of sketched update is the probabilistic quantization approach \cite{han2015deep}, in which the update matrices are vectorized and quantized for each scalar. To reduce the error from quantization, a structured random rotation that is the product of a Walsh-Hadamard matrix and
binary diagonal matrix \cite{suresh2017distributed} can be applied before quantization. 

The simulation results on the CIFAR-10 image classification task show that for structured updates, the random mask performs better than that of the low rank approach. The random mask approach also achieves higher accuracy than sketching approaches since the latter involves a removal of some information obtained during training. However, the combination of all three sketching tools, i.e., subsampling, quantization, and rotation, can achieve higher compression rate and faster convergence, albeit with some sacrifices in accuracy. For example, by using 2 bits for quantization and sketching out all but 6.25\% of update data, the number of bits needed to represent updates can be reduced by 256 times and the accuracy level achieved is 85\%. In addition, sketching updates can achieve higher accuracy in training when there are more participants trained per round. This suggests that for practical implementation of FL where there are many participants available, more participants can be selected for training per round so that subsampling can be more aggressive to reduce communication costs. 

\begin{figure}[!]
	\centering
	\includegraphics[width=\linewidth]{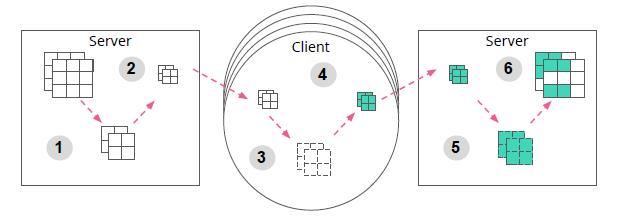}
	\caption{\small The compression techniques considered are summarized above by the diagram from authors in \cite{caldas2018expanding}. (i) Federated dropout to reduce size of model (ii) Lossy compression of model (iii) Decompression for training (iv) Compression of participant updates (v) Decompression (vi) Global aggregation}
	\label{fig:compress}
\end{figure}

The authors in \cite{caldas2018expanding} extend the studies in \cite{konevcny2016federated} by proposing lossy compression and federated dropout to reduce \textit{server-to-participant} communication costs. A summary of the proposed techniques are adapted from the authors' work in Fig. \ref{fig:compress}. For participant-to-server upload of model parameters that we discuss previously, the decompressions can be averaged over many updates to receive an unbiased estimate. However, there is no averaging for server-to-participant communications since the same global model is sent to all participants during each round of communication. Similar to \cite{konevcny2016federated}, subsampling and probabilistic quantization are considered. For the application of structured random rotation before subsampling and quantization, Kashin's representation \cite{kashin1977diameters} is applied instead of the Hadamard transformation since the former is found to perform better in terms of accuracy-size tradeoff. 

In addition to the subsampling and quantization approaches, the federated dropout approach is also considered in which a fixed number of activation functions at each fully-connected layer is removed to derive a smaller sub-model. The sub-model is then sent to the participants for training. The updated submodel can then be mapped back to the global model to derive a complete DNN model with all weights updated during subsequent aggregation. This approach reduces the server-to-participant communication cost, and also the size of participant-to-server updates. In addition, local computation is reduced since fewer parameters have to be updated. The simulations are performed on MNIST, CIFAR-10, and EMNIST \cite{cohen2017emnist} datasets. For the lossy compression, it is shown that the subsampling approach taken by \cite{konevcny2016federated} does not reach an acceptable level of performance. The reason is that the update errors can be averaged out for participant-to-server uploads but not for server-to-participant downloads. On the other hand, quantization with Kashin's representation can achieve the same performance as the baseline without compression while having communication cost reduced by nearly 8 times when the model is quantized to 4 bits. For federated dropout approaches, the results show that a dropout rate of 25\% of weight matrices of fully-connected layers (or filters in the case of CNN) can achieve acceptable accuracy in most cases while ensuring around 43\% reduction in size of models communicated. However, if dropout rates are more aggressive, convergence of the model can be slower.  

The aforementioned two studies suggest useful model compression approaches in reducing communication costs for both server-to-participant and participant-to-server communications. As one may expect, the reduction in communication costs come with sacrifices in model accuracy. It will thus be useful to formalize the compression-accuracy tradeoffs, especially since this varies for different tasks, or when different number of FL participants are involved.

%====================================================================
%====================================================================

\subsection{Importance-based Updating}

Based on the observation that most parameter values of a DNN model are sparsely distributed and close to zero \cite{strom2015scalable}, the authors in \cite{tao2018esgd} propose the edge Stochastic Gradient Descent (eSGD) algorithm that selects only a small fraction of important gradients to be communicated to the FL server for parameter update during each communication round. The eSGD algorithm keeps track of loss values at two consecutive training iterations. If the loss value of the current iteration is smaller than the preceding iteration, this implies that current training gradients and model parameters are important for training loss minimalization and thus, their respective hidden weights are assigned a positive value. In addition, the gradient is also communicated to the server for parameter update. Once this does not hold, i.e., the loss increases as compared to the previous iteration, other parameters are selected to be updated based on their hidden weight values. A parameter with larger hidden weight value is more likely to be selected since it has been labeled as important several times during training. To account for small gradient values that can delay convergence if they are ignored and not updated completely \cite{chen2016revisiting}, these gradient values are accumulated as residual values. Since the residuals may arise from different training iterations, each update to the residual is weighted with a discount factor using the momentum correction technique \cite{lin2017deep}. Once the accumulated residual gradient reaches a threshold, they are chosen to replace the least important gradient coordinates according to the hidden weight values. The simulation results show that eSGD with a 50\% drop ratio can achieve higher accuracy than that of the thresholdSGD algorithm proposed in \cite{strom2015scalable}, which uses a fixed threshold value to determine which gradient coordinates to drop. eSGD can also save a large proportion of gradient size communicated. However, eSGD still suffers from accuracy loss as compared to standard SGD approaches. For example, when tested on simple classification tasks using the MNIST dataset, the model accuracy converges to just 91.22\% whereas standard SGD can achieve 99.77\% accuracy. If extended to more sophisticated tasks, the accuracy can potentially deteriorate to a larger extent. In addition, the accuracy and convergence speed of the eSGD approach fluctuates arbitrarily based on hyperparameters used, e.g., minibatch size. As such, further studies have to be conducted to formally balance the tradeoffs between communication costs and training performance.

While \cite{tao2018esgd} studies the selective communication of gradients, the authors in \cite{wangcmfl} propose the Communication-Mitigated Federated Learning (CMFL) algorithm that uploads only relevant local model updates to reduce communication costs while guaranteeing global convergence. In each iteration, a participant's local update is first compared with the global update to identify if the update is relevant. A relevance score is computed where the score equates to percentage of same-sign parameters in the local and global update. In fact, the global update is not known in advance before aggregation. As such, the global update made in the previous iteration is used as an estimate for comparison since it was found empirically that more than 99\% of the normalized difference of two sequential global updates are smaller than 0.05 in both MNIST CNN and Next-Word-Prediction LSTM. An update is considered to be irrelevant if its relevance score is smaller than a predefined threshold. The simulation results show that CMFL requires 3.47 times and 13.97 times fewer communication rounds to reach 80\% accuracy for MNIST CNN and Next-Word-Prediction LSTM, respectively, as compared to the benchmark \textit{FedAvg} algorithm. In addition, CMFL can save significantly more communication rounds as compared to Gaia. Note that Gaia is a geo-distributed ML approach suggested in \cite{hsieh2017gaia} which measures relevance based on \textit{magnitude} of updates rather than sign of parameters. When applied with the aforementioned MOCHA algorithm \ref{stats} \cite{smith2017federated}, CMFL can reduce communication rounds by 5.7 times for the Human Activity Recognition dataset \cite{anguita2013public} and 3.3 times for the Semeion Handwritten Digit dataset \cite{buscema2009semeion}. In addition, CMFL can achieve slightly higher accuracy since it involves the elimination of irrelevant updates that are outliers which harm training.

\begin{table*}[]
\centering \caption{\small Approaches to communication cost reduction in FL. \label{tab:communication}}
\begin{adjustbox}{width=\textwidth}
\begin{tabular}{|l|l|l|l|}
\hline
\rowcolor[HTML]{C0C0C0} 
\multicolumn{1}{|c|}{\cellcolor[HTML]{C0C0C0}\textbf{Approaches}}                      & \multicolumn{1}{c|}{\cellcolor[HTML]{C0C0C0}\textbf{Ref.}} & \multicolumn{1}{c|}{\cellcolor[HTML]{C0C0C0}\textbf{Key Ideas}}                                                                                                                          & \multicolumn{1}{c|}{\cellcolor[HTML]{C0C0C0}\textbf{Tradeoffs and Shortcomings}}                                                     \\ \hline
                                                                                       & \cite{mcmahan2016communication}           & \begin{tabular}[c]{@{}l@{}}More local updates in between communication for global aggregation, to reduce \\ instances of communication\end{tabular}                                      & \begin{tabular}[c]{@{}l@{}}Increased computation cost and poor  performance\\ in non-IID setting\end{tabular}                        \\ \cline{2-4} 
                                                                                       & \cite{liu2019communication}               & \begin{tabular}[c]{@{}l@{}}Similar to the ideas of \cite{mcmahan2016communication}, but with convergence guarantees for vertical FL\end{tabular}                     & \begin{tabular}[c]{@{}l@{}}Increased computation cost and delayed\\ convergence if global aggregation is too infrequent\end{tabular} \\ \cline{2-4} 
                                                                                       & \cite{yao2018two}                         & \begin{tabular}[c]{@{}l@{}}Transfer learning-inspired two-stream model for FL participants to learn from the fixed \\global model so as to accelerate training convergence\end{tabular} & \begin{tabular}[c]{@{}l@{}}Increased computation cost and delayed \\ convergence\end{tabular}                                        \\ \cline{2-4} 
\multirow{-4}{*}{\begin{tabular}[c]{@{}l@{}}Edge and End \\ Computation\end{tabular}}  & \cite{liu2019edge}                        & \begin{tabular}[c]{@{}l@{}}MEC-inspired edge server assisted FL that aids in intermediate parameter aggregation to \\ reduce instances of communication\end{tabular}                     & \begin{tabular}[c]{@{}l@{}}System model is not scalable when there are \\ more edge servers\end{tabular}                             \\ \hline
                                                                                       & \cite{konevcny2016federated}              & \begin{tabular}[c]{@{}l@{}}Structured and sketched updates to compress local models communicated from participant\\  to FL server\end{tabular}                                           & Model accuracy and convergence issues                                                                                                \\ \cline{2-4} 
\multirow{-2}{*}{\begin{tabular}[c]{@{}l@{}}Model \\ Compression\end{tabular}}         & \cite{caldas2018expanding}                & \begin{tabular}[c]{@{}l@{}}Similar to the ideas of \cite{konevcny2016federated}, but for communication from  FL server \\  to participants\end{tabular}                   & Model accuracy and convergence issues                                                                                                \\ \hline
                                                                                       & \cite{tao2018esgd}                        & \begin{tabular}[c]{@{}l@{}}Selective communication of gradients that are assigned importance scores, \\ i.e., to reduce training loss\end{tabular}                                & \begin{tabular}[c]{@{}l@{}}Only empirically tested on simple datasets and \\ tasks, with fluctuating results\end{tabular}            \\ \cline{2-4} 
\multirow{-2}{*}{\begin{tabular}[c]{@{}l@{}}Importance-based \\ Updating\end{tabular}} & \cite{wangcmfl}                           & \begin{tabular}[c]{@{}l@{}}Selective communication of local model updates that have higher relevance scores when \\ compared to previous global model\end{tabular}                             & \begin{tabular}[c]{@{}l@{}}Difficult to implement when global aggregations\\ are less frequent\end{tabular}                          \\ \hline
\end{tabular}
\end{adjustbox}
\end{table*}

\subsection{Summary and Lessons Learned} 

In this section, we have reviewed three main approaches for communication cost reduction in FL, and for each approach, we discuss the solutions proposed in different studies. We summarize the approaches along with references in Table \ref{tab:communication}. From this review, we gather the following lessons learned:

\begin{itemize}
\item Communication cost is a key issue to be resolved before we can implement FL at scale. In particular, the state-of-the-art DL models have high inference accuracy but are increasingly complex with millions of parameters. The slow upload speed of mobile devices can thus impede the implementation of efficient FL. 

\item This section explores several key approaches to communication cost reduction. However, many of the approaches, e.g., model compression, result in a deterioration in model accuracy or incur high computation cost. For example, when too many local updates are implemented between communication rounds, the communication cost is indeed reduced but the convergence can be significantly delayed \cite{liu2019communication}. The tradeoff between these sacrifices and communication cost reduction thus has to be well-managed. 

\item The current studies of this tradeoff are often mainly empirical in nature, e.g., several experiments have to be done to find the optimal number of local training iterations before communication. With more effective optimization approaches formalized theoretically and tested empirically, FL can eventually become more scalable in nature. For example, the authors in \cite{yang2019energy} study the tradeoffs between the completion time of FL training and energy cost expended. Then, a weighted sum of completion time and energy consumption is minimized using an iterative algorithm. For delay-sensitive scenarios, the weights can be adjusted such that the FL participants consume more energy for completion time minimization.

\item Apart from working to directly reduce the size of model communicated, studies on FL can draw inspiration from applications and approaches in the MEC paradigm. For example, a simple case study introduced in \cite{tao2018esgd} considers the base station as an intermediate model aggregator to reduce instances of device-cloud communication. Unfortunately, there are convergence issues when more edge servers or mobile devices are considered. This is exacerbated by the non-IID distribution of data across different edge nodes. For future works, this statistical challenge can be met, e.g., through inspirations from multi-task learning as we have discussed in Section \ref{stats}. In addition, more effective and innovative system models can be explored such that FL networks can utilize the wealth of computing and storage resources that are closer to the data sources to facilitate efficient FL.

\item For the studies that we have discussed in this section, the heterogeneity among mobile devices, e.g., in computing capabilities, is often not considered. For example, one of the ways to reduce communication cost is to increase computation on edge devices, e.g., by performing more local updates \cite{mcmahan2016communication} before each communication round. In fact, this does not merely lead to the expenditure of greater computation cost. The approach may also not be feasible for devices with weak processing power, and can lead to the \textit{straggler effect}. As such, we further explore issues on resource allocation in the next section.

\end{itemize}

%====================================================================
%====================================================================

\section{Resource Allocation}
\label{sec:resource}
%\cite{valerio2017communication} 
%\cite{wang2019adaptive} \cite{sprague2018asynchronous} \cite{nishio2018client} \cite{hamidouche2018collaborative} \cite{lin2017deep} \cite{samarakoon2018distributed} \cite{chen2018federated} \cite{li2018fog} \cite{wang2018edge}
%\cite{feng2018joint} \cite{li2018learning} \cite{amiri2019machine}  \cite{liu2018wireless} \cite{li2018edge} \cite{anh2018efficient} 
%\cite{valerio2018energy}

%====================================================================
%====================================================================

FL involves the participation of heterogeneous devices that have different dataset qualities, computation capabilities, energy states, and willingness to participate. Given the device heterogeneity and resource constraints, i.e., in device energy states and communication bandwidth, resource allocation has to be optimized to maximize the efficiency of the training process. In particular, the following resource allocation issues need to be considered:

\begin{itemize}

\item \textit{Participant Selection:} As part of the FL protocol presented in Section \ref{proto}, participant selection refers to the selection of devices to participate in each training round. Typically, a set of participants is randomly selected by the server to participate. Then, the server has to aggregate parameter updates from all participating devices in the round before taking a weighted average of the models \cite{mcmahan2016communication}. As such, the training progress of FL is limited by the training time of the slowest participating devices, i.e., stragglers \cite{sprague2018asynchronous}. New participant selection protocols are thus investigated in order to address the training bottleneck in FL. 

\item \textit{Joint Radio and Computation Resource Management:} Even though computation capabilities of mobile devices have grown rapidly, many devices still face a scarcity of radio resources \cite{jordan2019communication}. Given that local model transmission is an integral part of FL, there has been a growing number of studies that focus on developing novel wireless communication techniques for efficient FL.

\item \textit{Adaptive Aggregation:} As discussed in Section \ref{fltrain}, FL involves global aggregation in which model parameters are communicated to the FL server for aggregation. The conventional approach to global aggregation is a synchronous one, i.e., global aggregations occur in fixed intervals after all participants complete a certain number of rounds of local computation. However, adaptive calibrations of global aggregation frequency can be investigated to increase training efficiency subject to resource constraints \cite{sprague2018asynchronous}.

\item \textit{Incentive Mechanism:}  In the practical implementation of FL, participants may be reluctant to participate in a federation without receiving compensation since training models is resource-consuming. In addition, there exists information asymmetry between the FL server and participants since participants have greater knowledge of their available computation resources and data quality. Therefore, incentive mechanisms have to be carefully designed to both incentivize participation and reduce the potential adverse impacts of information asymmetry.

 \end{itemize}

\subsection{Participant Selection}

\begin{figure}[!]
	\centering 
	\includegraphics[width=\columnwidth]{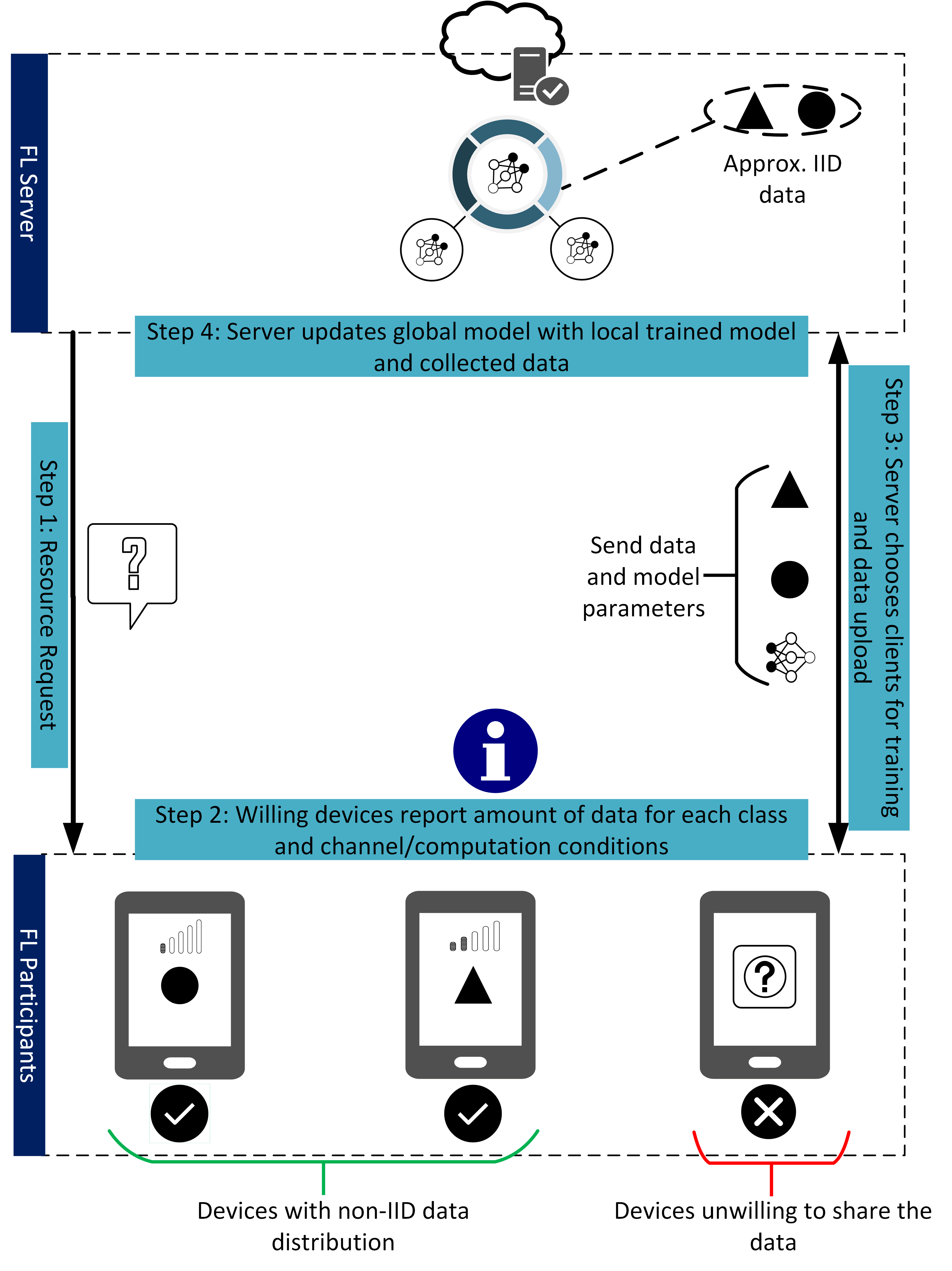}
	\caption{\small Participant selection under the FedCS and Hybrid-FL protocol.}
	\label{fig:fedcs}
\end{figure}

To mitigate the training bottleneck, the authors in \cite{nishio2018client} propose a new FL protocol called \textit{FedCS}. This protocol is illustrated  in Fig. \ref{fig:fedcs}. The system model is a MEC framework in which the operator of the MEC is the FL server that coordinates training in a cellular network that comprises participating mobile devices that have heterogeneous resources. Accordingly, the FL server first conducts a \textit{Resource Request} step to gather information such as wireless channel states and computing capabilities from a subset of randomly selected participants. Based on this information, the MEC operator selects the maximum possible number of participants that can complete the training within a prespecified deadline for the subsequent global aggregation phase. By selecting the maximum possible number of participants in each round, accuracy and efficiency of training are preserved. To solve the maximization problem, a greedy algorithm \cite{sviridenko2004note} is proposed, i.e., participants that take the least time for model upload and update are iteratively selected for training. The simulation results show that compared with the FL protocol which only accounts for training deadline without performing participant selection, \textit{FedCS} can achieve higher accuracy since \textit{FedCS} is able to involve more participants in each training round \cite{mcmahan2016communication}. However, \textit{FedCS} has been tested only on simple DNN models. When extended to the training of more complex models, it may be difficult to estimate how many participants should be selected. For example, more training rounds may be needed for the training of complex models, and the selection of too few participants may lead to poor performance considering that some participants may drop out during training. In addition, there is bias towards selecting participants with devices that have better computing capabilities. These participants may not hold data that is representative of the population distribution. In particular, we revisit the fairness issue \cite{mohri2019agnostic} subsequently in this section. 

While \textit{FedCS} addresses heterogeneity of resources among participants in FL, the authors in \cite{yoshida2019hybrid} extend their work on the \textit{FedCS} protocol with the Hybrid-FL protocol that deals with differences in data distributions among participants. The dataset of participants participating in FL may be non-IID since it is reflective of each individual user's specific characteristics. As we have discussed in Section \ref{stats}, the non-IID dataset may significantly degrade the performance of the \textit{FedAvg} algorithm \cite{zhao2018federated}. One proposed measure to address the non-IID nature of the dataset is to distribute publicly available data to participants, such that the EMD between their on-device dataset and the population distance is reduced. However, such a dataset may not always exist, and participants may not download them for security reasons. Thus, an alternative solution is to construct an approximately IID dataset using inputs from a limited number of privacy insensitive participants \cite{yoshida2019hybrid}. In the Hybrid-FL protocol, during the \textit{Resource Request} step  (Fig.~\ref{fig:fedcs}), the MEC operator asks random participants if they permit their data to be uploaded. During the participant selection phase, apart from selecting participants based on computing capabilities, participants are selected such that their uploaded data can form an approximately IID dataset in the server, i.e., the amount of collected data in each class has close values (Fig.~\ref{fig:fedcs}). Thereafter, the server trains a model on the collected IID dataset and merge this model with the global model trained by participants. The simulation results show that even with just 1\% of participants sharing their data, classification accuracy for non-IID data can be significantly improved as compared to the aforementioned \textit{FedCS} benchmark where data is not uploaded at all. However, the recommended protocol can violate the privacy and security of users, especially if the FL server is malicious. In the case when participants are malicious, data can be falsified before uploading, as we will further discuss in Section \ref{sec:security}. In addition, the proposed measure can be costly especially in the case of videos and images. As such, it is unlikely that participants will volunteer for data uploading when they can free ride on the efforts of other volunteers. For feasibility, a well-designed incentive and reputation mechanism is needed to ensure that only trustworthy participants are allowed to upload their data. 

In general, the mobile edge network environment in which FL is implemented on is dynamic and uncertain with variable constraints, e.g., wireless network and energy conditions. Thus, this can lead to training bottlenecks. To this end, Deep Q-Learning (DQL) can be used to optimize resource allocation for model training as proposed in \cite{anh2018efficient}. The system model is a Mobile Crowd Machine Learning (MCML) setting which enables participants in a mobile crowd network to collaboratively train DNN models required by a FL server. The participating mobile devices are constrained by energy, CPU, and wireless bandwidth. Thus, the server needs to determine proper amounts of data, energy, and CPU resources that the mobile devices use for training to minimize energy consumption and training time. Under the uncertainty of the mobile environment, a stochastic optimization problem is formulated. In the problem, the server is the agent, the state space includes the CPU and energy states of the mobile devices, and the action space includes the number of data units and energy units taken from the mobile devices. To achieve the objective, the reward function is defined as a function of the accumulated data, energy consumption, and training latency. To overcome the large state and action space issues of the server, the DQL technique based on Double Deep Q-Network (DDQN) \cite{van2016deep} is adopted to solve the server's problem. The simulation results show that the DQL scheme can reduce energy consumption by around 31\% compared with the greedy algorithm, and training latency is reduced up to 55\% compared with the random scheme. However, the proposed scheme is applicable only in federations with few participating mobile devices. 

As an extension to \cite{anh2018efficient}, the authors in \cite{nguyen2019resource} also proposes a resource allocation approach using DRL, with the added uncertainty that FL participants are mobile and so they may venture out of the network coverage range with a certain probability. Without prior knowledge of the mobile network, the FL server is able to optimize resource allocation across participants, e.g., channel selection and device energy consumption.

The aforementioned resource allocation approaches focus on improving the training efficiency of FL. However, this may cause some FL participants to be left out of the aggregation phase because they are stragglers with limited computing or communication resources. 

One consequence of this  \textit{unfair} resource allocation, a topic that is commonly explored in resource allocation for wireless networks \cite{neely2008fairness} and ML \cite{feldman2015computational}. For example, if the participant selection protocol selects mobile devices with higher computing capabilities to participate in each training round \cite{nishio2018client}, the FL model will be overrepresented by the distribution of data owned by participants with devices that have higher computing capabilities. Therefore, the authors in \cite{mohri2019agnostic} and \cite{li2019fair} consider fairness as an additional objective in FL. Fairness is defined in \cite{li2019fair} to be the \textit{variance} of performance of an FL model across participants. If the variance of the testing accuracy is large, this implies the presence of more bias or less fairness, since the learned model may be highly accurate for certain participants and less so for other underrepresented participants. The authors in \cite{li2019fair} propose the \textit{q}-Fair FL (\textit{q}-FFL) algorithm that reweighs the objective function in \textit{FedAvg} to assign higher weights in the loss function to devices with higher loss. The modified objective function is as follows:
\begin{equation}
\min _{w} F_{q}(w)=\sum_{k=1}^{m} \frac{p_{k}}{q+1} F_{k}^{q+1}(w)
\end{equation}
where $F_k$ refers to the standard loss functions presented in Table \ref{tab:loss-functions}, $q$ refers to the calibration of fairness in the system model, i.e., setting $q=0$ returns the formulation to the typical FL objective, and $p_{k}$ refers to ratio of local samples to the total number of training samples. In fact, this is a generalization of the Agnostic FL (AFL) algorithm proposed in \cite{mohri2019agnostic}, in which the device with the \text{highest} loss dominates the entire loss function. The simulation results show that the proposed \textit{q}-FFL can achieve lower variance of testing accuracy and converges more quickly than the AFL algorithm. However, as expected, for some calibrations of the \textit{q}-FFL algorithm, there can be convergence slowdown since stragglers can delay the training process. As such, an asynchronous aggregation approach (to be subsequently discussed in this section) can potentially be considered for use with the \textit{q}-FFL algorithm.

In contrast, the authors in \cite{chen2020convergence} propose a neural network based approach to estimate the local models of FL participants that are left out during training. In the system model, resource blocks are first allocated by the base station to users whose models have larger effects on the global FL model. In particular, one user is selected to always be connected to the base station. This user's model parameters are then used as input to the feedforward neural network to estimate the model parameters of users who are left out during the training iteration. This allows the base stations to be able to integrate more locally trained FL model parameters to each iteration of global aggregation, thus improving the FL convergence speed.

\subsection{Joint Radio and Computation Resource Management}

\begin{figure}[!]
	\centering
	\includegraphics[width=\linewidth, height= 9cm]{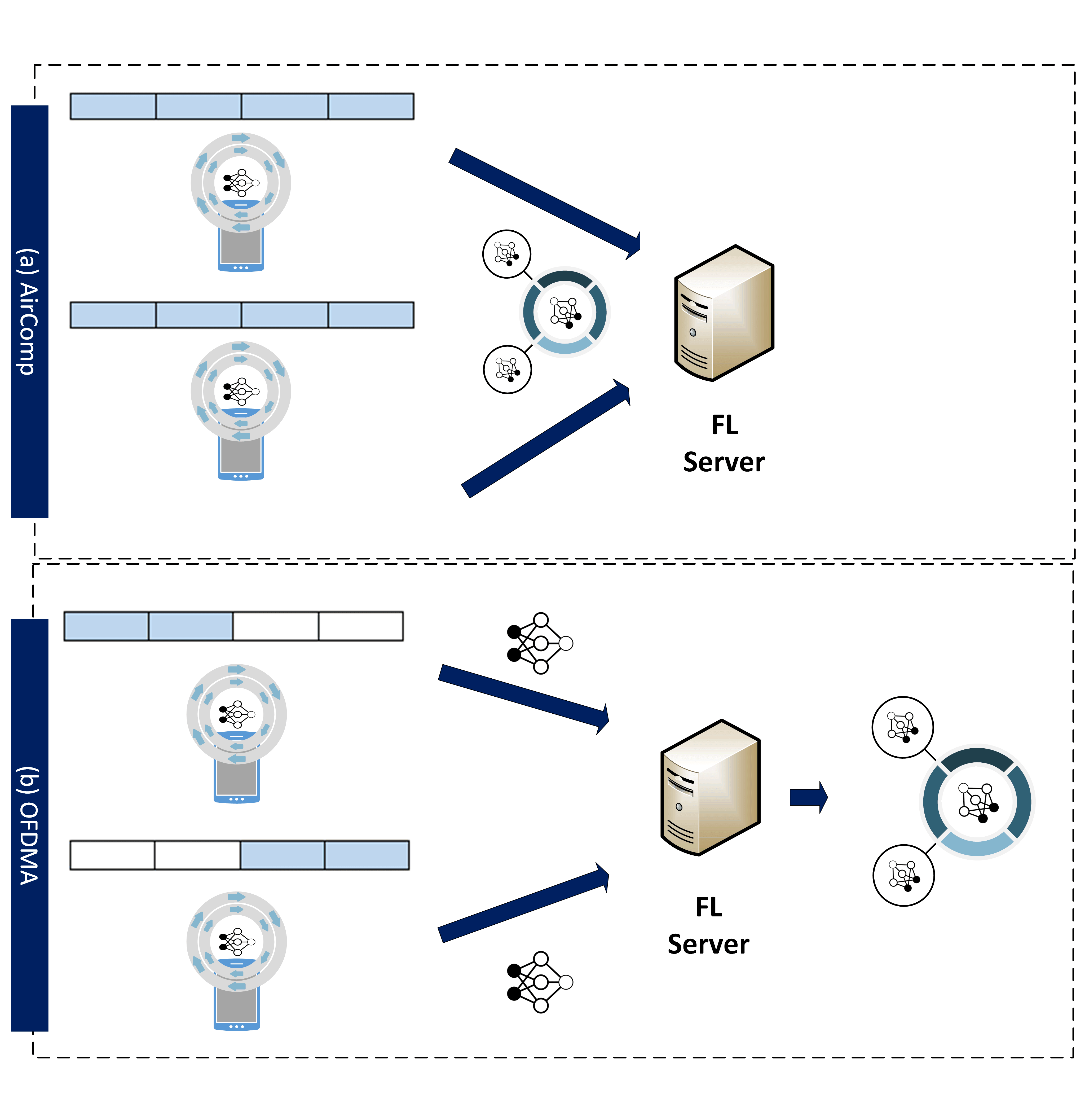}
	\caption{\small A comparison \cite{zhu2018low} between (a) BAA by over-the-air computation which reuses bandwidth (above) and (b) OFDMA (below) which uses only the allocated bandwidth.}
	\label{fig:aircomp}
\end{figure}

While most FL studies have previously assumed orthogonal-access schemes such as Orthogonal Frequency-division Multiple Access (OFDMA) \cite{lopez2009ofdma}, the authors in \cite{zhu2018low} propose a multi-access Broadband Analog Aggregation (BAA) design for communication-latency reduction in FL. Instead of performing communication and computation separately during global aggregation at the server, the BAA scheme builds on the concept of over-the-air computation \cite{nazer2011compute} to \textit{integrate} computation and communication through exploiting the signal superposition property of a multiple-access channel. The proposed BAA scheme allows the reuse of the whole bandwidth (Fig. \ref{fig:aircomp}(a)) whereas OFDMA orthogonalizes bandwidth allocation (Fig. \ref{fig:aircomp}(b)). As such, for orthogonal-access schemes, communication latency increases in direct proportion with the number of participants whereas for multi-access schemes, latency is independent of the number of participants. The bottleneck of signal-to-noise ratio (SNR) during BAA transmission is the participating device with the longest propagation distance given that devices that are nearer have to lower their transmission power for amplitude alignment with devices located further. To increase SNR, participants with longer propagation distance have to be dropped. However, this leads to the truncation of model parameters. As such, to manage the SNR-truncation tradeoff, three scheduling schemes are considered namely i) \textit{Cell-interior scheduling}: participants beyond a distance threshold are not scheduled, ii) \textit{All-inclusive scheduling}: all participants are considered, and iii) \textit{Alternating scheduling}: edge server alternates between the two aforementioned schemes. The simulation results show that the proposed BAA scheme can achieve similar test accuracy as the OFDMA scheme while achieving latency reduction from 10 times to 1000 times. As a comparison between the three scheduling schemes, the cell-interior scheme outperforms the all-inclusive scheme in terms of test accuracy for high mobility networks where participants have rapidly changing locations. For low mobility networks, the alternating scheduling scheme outperforms cell-interior scheduling. 

As an extension, the authors in \cite{amiri2019federated} also introduce error accumulation and gradient sparsification in addition to over-the-air computation. In \cite{zhu2018low}, gradient vectors that are not transmitted as a result of power constraints are completely dropped. To improve the model accuracy, the untransmitted gradient vectors can first be stored in an error accumulation vector. In the next round, local gradient estimates are then corrected using the error vector. In addition, when there are bandwidth limitations, the participating device can apply gradient sparsification to keep only elements with the highest magnitudes for transmission. The elements that are not transmitted are subsequently added on to the error accumulation vector for gradient estimate correction in the next round. The simulation results show that the proposed scheme can achieve higher test accuracy than over-the-air computation without error accumulation or gradient sparsification since it corrects gradient estimates with the error accumulation vector and allows for a more efficient utilization of the bandwidth. 

Similar to \cite{zhu2018low} and \cite{amiri2019federated}, the authors in \cite{yang2018federated} propose an integration of computation and communication via over-the-air computation. However, it is observed that aggregation error incurred during over-the-air computation can lead to a drop in model accuracy \cite{keskar2016large} as a result of signal distortion. As such, a participant selection algorithm is proposed in which the number of devices selected for training is maximized so as to improve statistical learning performance \cite{mcmahan2016communication} while keeping the signal distortion below a threshold. Due to the nonconvexity \cite{boyd2004convex} of the mean-square-error (MSE) constraint and intractability of the optimization problem, a difference-of-convex functions (DC) algorithm \cite{tao2005dc} is proposed to solve the maximization problem. The simulation results show that the proposed DC algorithm is scalable and can also achieve near-optimal performance that is comparable to global optimization, which is non-scalable due to its exponential time complexity. In comparison with other state-of-the-art approaches such as the semidefinite relaxation technique (SDR) proposed in \cite{luo2007approximation}, the proposed DC algorithm can also select more participants, thus also achieving higher model accuracy.

%====================================================================
%====================================================================

%====================================================================
%====================================================================

\subsection{Adaptive Aggregation}

\begin{figure}[!]
	\centering
	\includegraphics[width=\columnwidth]{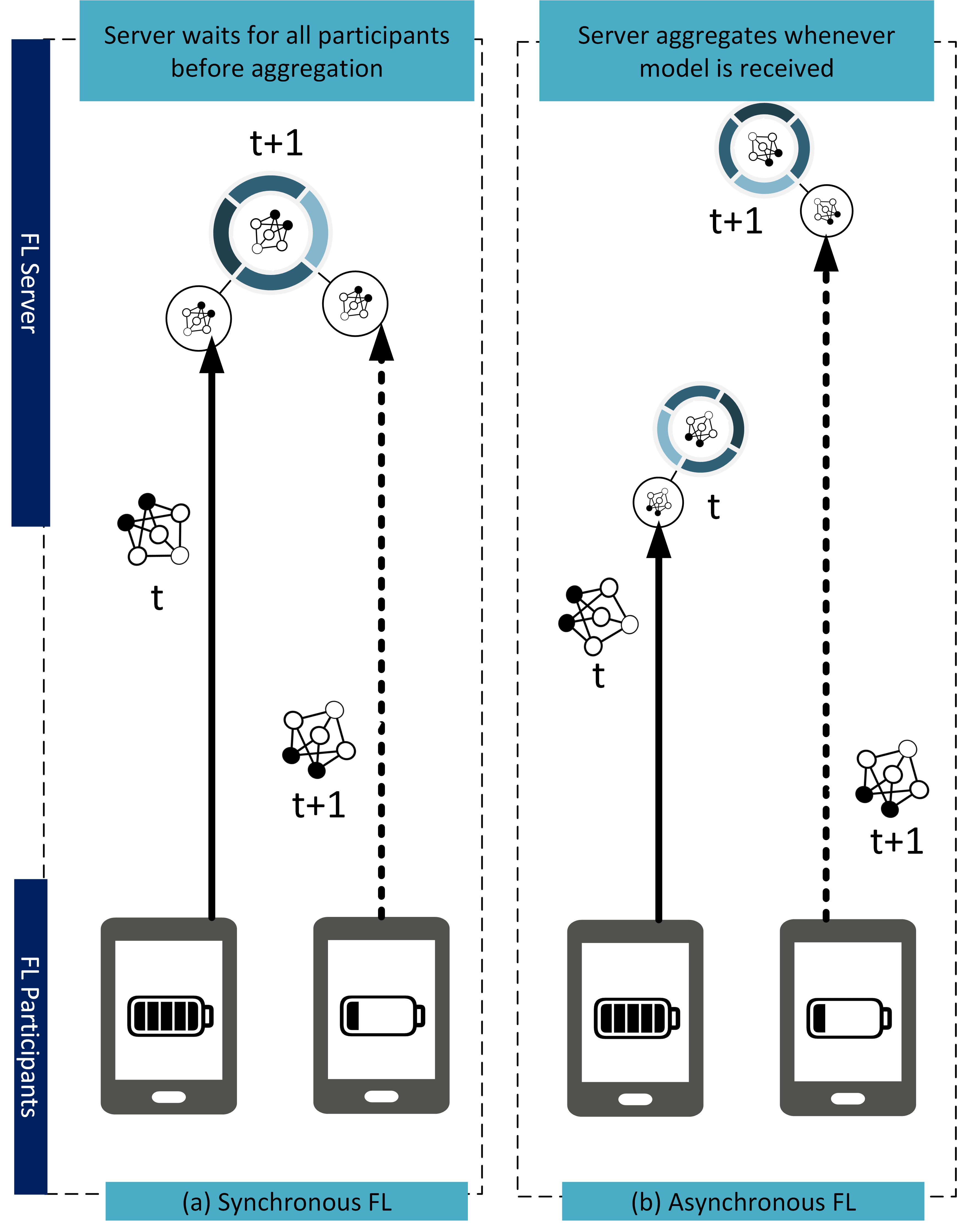}
	\caption{\small A comparison between (a) synchronous and (b) asynchronous FL.}
	\label{fig:async}
\end{figure}

The proposed \textit{FedAvg} algorithm synchronously aggregates parameters as shown in Fig. \ref{fig:async}(a) and is thus susceptible to the straggler effect, i.e., each training round only progresses as fast as the slowest device since the FL server waits for \textit{all} devices to complete local training before global aggregation can take place \cite{sprague2018asynchronous}. In addition, the model does not account for participants that can join halfway when the training round is already in progress. As such, the asynchronous model is proposed to improve the scalability and efficiency of FL. For asynchronous FL, the server updates the global model whenever it receives a local update (Fig. \ref{fig:async}(b)). The authors in \cite{sprague2018asynchronous} find empirically that an asynchronous approach is robust to participants joining halfway during a training round, as well as when the federation involves participating devices with heterogeneous processing capabilities. However, the model convergence is found to be significantly delayed when data is non-IID and unbalanced. As an improvement, \cite{xie2019asynchronous} propose the \textit{FedAsync} algorithm in which each newly received local updates are adaptively weighted according to staleness, that is defined as the difference between the current epoch and iteration in which the received update belongs to. For example, a stale update from a straggler is outdated since it should have been received in previous training rounds. As such, it is weighted less. In addition, the authors also prove the convergence guarantee for a restricted family of non-convex problems. However, the current hyperparameters of the \textit{FedAsync} algorithm still have to be tuned to ensure convergence in different settings. As such, the algorithm is still unable to generalize to suit the dynamic computation constraints of heterogeneous devices. In fact, given the uncertainty surrounding the reliability of asynchronous FL, synchronous FL remains to be the approach most commonly used today \cite{bonawitz2019towards}.

For most existing implementations of the \textit{FedAvg} algorithm, the global aggregation phase occurs after a fixed number of training rounds. To better manage the dynamic resource constraints, the authors in \cite{wang2019adaptive} propose an adaptive global aggregation scheme which varies the global aggregation frequency so as to ensure desirable model performance while ensuring an efficient use of available resources, e.g., energy, during the FL training process. In \cite{wang2019adaptive},  the MEC system model  used consists of (i) the local update phase where the model is trained using local data, (ii) edge aggregation phase where the intermediate aggregation occurs and (iii) global aggregation phase where updated model parameters are received and aggregated by the FL server. In particular, the authors study how the training loss is affected when the total number of edge server aggregation and local updates between global aggregation intervals vary. For this, a convergence bound of gradient descent with non-IID data is first derived. Then, a control algorithm is subsequently proposed to adaptively choose the optimal global aggregation frequency based on the most recent system state. For example, if global aggregation is too time consuming, more edge aggregations will take place before communication with the FL server is initiated. The simulation results show that the adaptive aggregation scheme outperforms the fixed aggregation scheme in terms of loss function minimization and accuracy within the same time budget. However, the convergence guarantee of the adaptive aggregation scheme is only considered for convex loss functions currently.

%====================================================================
%====================================================================

\subsection{Incentive Mechanism}
 
The authors in \cite{feng2018joint} propose a service pricing scheme in which participants serve as training service providers for a model owner. In addition, to overcome energy inefficiency in the transfer of model updates, a cooperative relay network is proposed to support model update transfer and trading. The interaction between participants and model owner is modelled as a Stackelberg game \cite{osborne2004introduction} in which the model owner is the buyer and participants are the sellers. The Stackelberg game is proposed in which each rational participant can noncooperatively decide on its own profit maximization price. In the lower-level subgame, the model owner determines size of training data to maximize profits with consideration of the increasing concave relationship between learning accuracy of the model and size of training data. In the upper-level subgame, the participants decide the price per unit of data to maximize their individual profits. The simulation results show that the proposed mechanism can ensure uniqueness of the Stackelberg equilibrium. For example, model updates that contain valuable information are priced higher at the Stackelberg equilibrium. In addition, model updates can be transferred cooperatively, thus reducing congestion in communication and improving energy efficiency. However, the simulation environment only involves relatively few mobile devices.

Similar to \cite{feng2018joint}, the authors in \cite{sarikaya2019motivating,khan2019federated} also model the interaction between participants and model owner as a Stackelberg game, which is well-suited to represent the FL server-participant interaction involved in FL. 

However, unlike the aforementioned conventional approaches to solving Stackelberg formulations, a DRL-based approach is adopted together with the Stackelberg game by the authors in \cite{zhan2020learning}. In the DRL formulation, the FL server acts as an agent that decides a payment in response to the participation level and payment history of edge nodes, with the objective of minimizing incentive expenses. Then, the edge nodes determine an optimal participation level in response to the payment policy. This learning based incentive mechanism design enables the FL server to derive an optimal policy in response to its observed state, without requiring any prior information.

 \begin{figure}[!]
	\centering
	\includegraphics[width=0.9\linewidth]{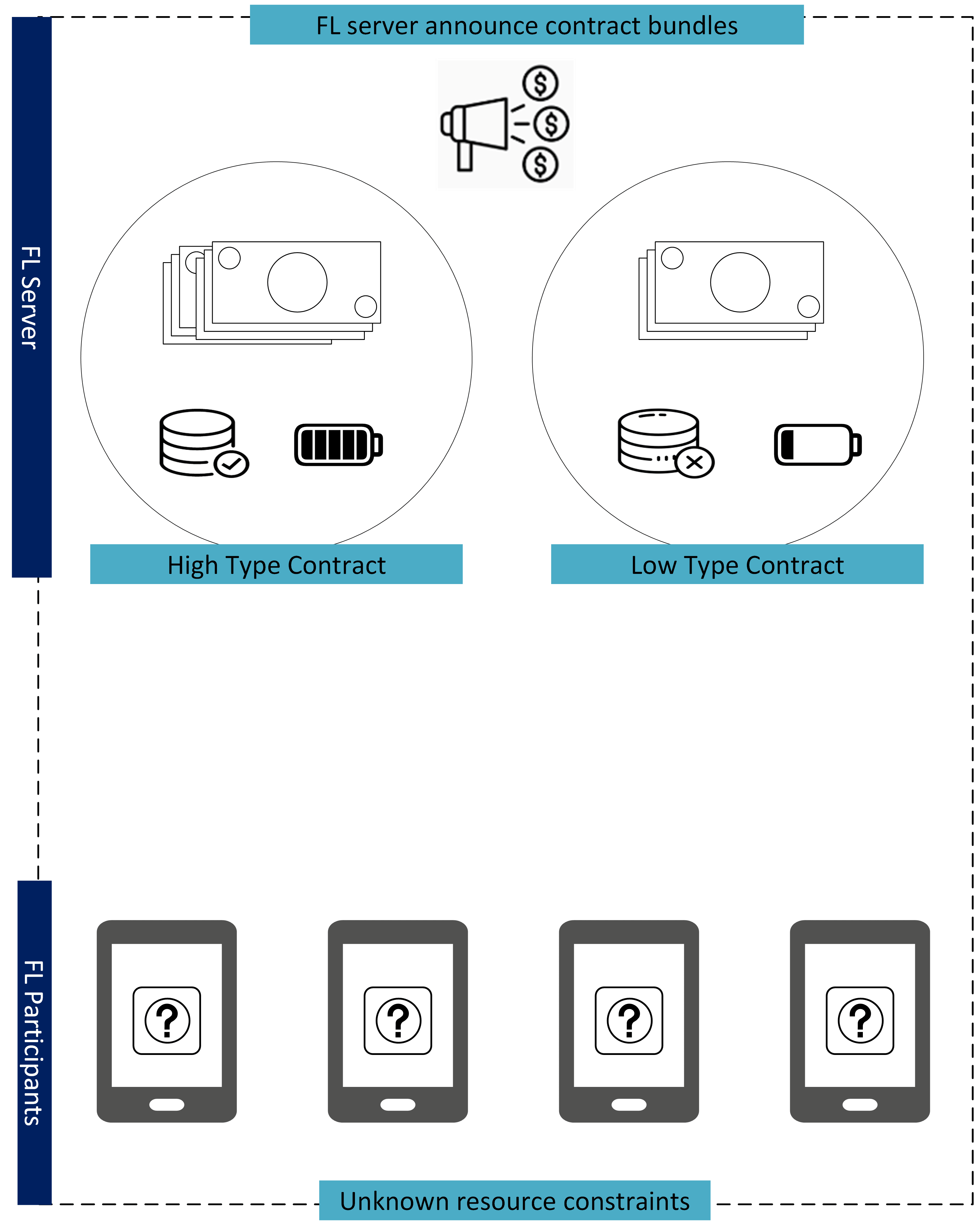}
	\caption{\small Participants with unknown resource constraints maximize their utility only if they choose the bundle that best reflects their constraints.}
	\label{fig:contract}
\end{figure}

In contrast to \cite{feng2018joint,sarikaya2019motivating,khan2019federated,zhan2020learning}, the authors in \cite{kang2019incentive} propose an incentive design using a contract theoretic \cite{bolton2005contract} approach to attract participants with high-quality data for FL. In particular, well-designed contracts can reduce information asymmetry through self-revealing mechanisms in which participants select only the contracts specifically designed for their types. For feasibility, each contract must satisfy the Individual Rationality (IR) and Incentive Compatibility (IC) constraints. For IR, each participant is assured of a positive utility when the participant participates in the federation. For IC, every utility maximizing participant only chooses the contract designed for its type. The model owner aims to maximize its own profits subject to IR and IC constraints. As illustrated in Fig. \ref{fig:contract}, the optimal contracts derived are self-revealing such that each high-type participant with higher data quality only chooses contracts designed for its type, whereas each low-type participant with lower data quality does not have the incentive to imitate high-type participants.  The simulation results show that all types of participants only achieve maximum utility when they choose the contract that matches their types. In addition, the proposed contract theory approach also has better performance in terms of profit for the model owner compared with the Stackelberg game-based incentive mechanism. This is because under the contract theoretic approach, the model owner can extract more profits from the participants whereas under the Stackelberg game approach, the participants can optimize their individual utilities. In fact, the information asymmetry between FL servers and participants make contract theory a powerful and efficient tool for mechanism design in FL. As an extension, the authors in \cite{ye2020federated} introduced a \textit{multi-dimensional} contract in which each FL participant determines the optimal computation power and image quality it is willing to contribute for model training, in exchange for contract rewards in each iteration.

The authors in \cite{kang2019incentive} further introduce reputation as a metric to measure the reliability of FL participants and design a reputation-based participant selection scheme for reliable FL \cite{kangincentive2}. In this setting, each participant has a reputation value \cite{jurca2003incentive} derived from two sources, (i) direct reputation opinions from past interactions with the FL server and (ii) indirect reputation opinions from other task publishers, i.e., other FL servers. The indirect reputation opinions are stored in an open-access reputation blockchain \cite{dennis2015rep} to ensure secure reputation management in a decentralized manner. Before model training, the participants choose a contract that best fits its dataset accuracy and resource conditions. Then, the FL server chooses the participants that have reputation scores which are larger than a prespecified threshold. After the FL task is completed, i.e., a desirable accuracy is achieved, the FL server updates the reputation opinions, which are subsequently stored in the reputation blockchain. The simulation results show that the proposed scheme can significantly improve the accuracy of the FL model since unreliable workers are detected and not selected for FL training.

\begin{table*}[]
\centering \caption{\small Approaches to resource allocation in FL. \label{tab:resource}}
\begin{adjustbox}{width=\textwidth}
\begin{tabular}{|l|l|l|l|}
\hline
\rowcolor[HTML]{C0C0C0} 
\multicolumn{1}{|c|}{\cellcolor[HTML]{C0C0C0}\textbf{Approaches}}                                              & \multicolumn{1}{c|}{\cellcolor[HTML]{C0C0C0}\textbf{Ref.}}                                      & \multicolumn{1}{c|}{\cellcolor[HTML]{C0C0C0}\textbf{Key Ideas}}                                                                                                                  & \multicolumn{1}{c|}{\cellcolor[HTML]{C0C0C0}\textbf{Tradeoffs and Shortcomings}}                                                                                                               \\ \hline
                                                                                                               & \cite{nishio2018client}                                                        & \begin{tabular}[c]{@{}l@{}}FedCS to select participants based on computation capabilities so as to\\ complete FL training before specified deadline\end{tabular}                 & \begin{tabular}[c]{@{}l@{}}Difficult to estimate training duration accurately for\\ complex models\end{tabular}                                                                                           \\ \cline{2-4} 
                                                                                                               & \cite{yoshida2019hybrid}                                                       & \begin{tabular}[c]{@{}l@{}}Following \cite{nishio2018client}, Hybrid-FL to select participants so as to accumulate \\ IID, distributable data for FL model training\end{tabular}           & \begin{tabular}[c]{@{}l@{}}Request of data sharing may defeat the \\ original intent of FL\end{tabular}                                                                                        \\ \cline{2-4} 
                                                                                                               & \cite{anh2018efficient}                                                        & DRL to determine resource consumption by FL participants                                                                                                                         &                                                                                                                                                                                                \\ \cline{2-3}
                                                                                                               & \cite{nguyen2019resource}                                                      & \begin{tabular}[c]{@{}l@{}}Following \cite{anh2018efficient}, DRL for resource allocation with\\ mobility-aware FL participants\end{tabular}                    & \multirow{-2}{*}{\begin{tabular}[c]{@{}l@{}}DRL models are difficult to train especially\\ when the number of FL participants are large\end{tabular}}                                          \\ \cline{2-4} 
\multirow{-5}{*}{\begin{tabular}[c]{@{}l@{}}Participant\\ Selection\end{tabular}}                              & \cite{li2019fair}                                                              & Fair resource allocation to reduce variance of model performance                                                                                                                 & Convergence delays with more fairness                                                                                                                                                                            \\ \hline
                                                                                                               & \cite{zhu2018low}                                                              & \begin{tabular}[c]{@{}l@{}}Propose BAA to integrate computation and communication through\\ exploiting the signal superposition property of multiple-access channel\end{tabular} &                                                                                                                                                                                                \\ \cline{2-3}
                                                                                                               & \cite{amiri2019federated}                                                      & \begin{tabular}[c]{@{}l@{}}Improves on \cite{zhu2018low} by accounting for gradient vectors that are not\\ transmitted due to power constraints\end{tabular}    &                                                                                                                                                                                                \\ \cline{2-3}
\multirow{-3}{*}{\begin{tabular}[c]{@{}l@{}}Joint Radio and \\ Computation\\ Resource Management\end{tabular}} & \cite{yang2018federated}                                                       & \begin{tabular}[c]{@{}l@{}}Improves on \cite{zhu2018low} using the DC algorithm to minimize\\ aggregation error\end{tabular}                                    & \multirow{-3}{*}{\begin{tabular}[c]{@{}l@{}}Signal distortion can lead to drop in accuracy,\\ the scalability is also an issue when large \\ heterogeneous networks are involved\end{tabular}} \\ \hline
                                                                                                               & \cite{sprague2018asynchronous}                                                 & \begin{tabular}[c]{@{}l@{}}Asynchronous FL where model aggregation occurs whenever local\\ updates are received by FL server\end{tabular}                                        & \begin{tabular}[c]{@{}l@{}}Significant delay in convergence in non-IID\\ and unbalanced dataset\end{tabular}                                                                                   \\ \cline{2-4} 
\multirow{-2}{*}{\begin{tabular}[c]{@{}l@{}}Adaptive \\ Aggregation\end{tabular}}                              & \cite{wang2019adaptive}                                                        & Adaptive global aggregation frequency based on resource constraints                                                                                                                                            & \begin{tabular}[c]{@{}l@{}}Convergence guarantees are limited to restrictive\\ assumptions\end{tabular}                                                                                        \\ \hline
                                                                                                               & \cite{feng2018joint,sarikaya2019motivating,khan2019federated,zhan2020learning} & \begin{tabular}[c]{@{}l@{}}Stackelberg game for incentivizing higher quantities of training data\\ or compute resource contributed \end{tabular}                                              & \begin{tabular}[c]{@{}l@{}}FL server derives lower profits. Also, assumption\\ that there is only one FL server\end{tabular}                                                                   \\ \cline{2-4} 
                                                                                                               & \cite{kang2019incentive,ye2020federated}                                       & Contract theoretic approach to incentivize FL participants                                                                                                                                                                  &                                                                                                                                                                                                \\ \cline{2-3}
\multirow{-3}{*}{\begin{tabular}[c]{@{}l@{}}Incentive \\ Mechanism\end{tabular}}                               & \cite{kangincentive2}                                                          & Reputation mechanism to select effective workers                                                                                                                                                             & \multirow{-2}{*}{Assumption that there is only one FL server}                                                                                                                                  \\ \hline
\end{tabular}
\end{adjustbox}
\end{table*}

\subsection{Summary and Lessons Learned} 

In this section, we have discussed four main issues in resource allocation. The issues and approaches are summarized in Table \ref{tab:resource}. From this section, the lessons learned are as follows:

\begin{itemize}

\item In heterogeneous mobile networks, the consideration of resource allocation is important to ensure efficient FL. For example, each training iteration is only conducted as quickly as the slowest FL participant, i.e., the straggler effect. In addition, the model accuracy is highly dependent on the quality of data used for training by FL participants. In this section, we have explored different dimensions of resource heterogeneity for consideration, e.g., varying computation and communication capabilities, willingness to participate, and quality of data for local model training. In addition, we have explored various tools that can be considered for resource allocation. For example, DRL is useful given the dynamic and uncertain wireless network conditions experienced by FL participants, whereas contract theory can serve as a powerful tool in mechanism design under the context of information asymmetry. Naturally, traditional optimization approaches have also been well explored in radio resource management for FL, given the high dependency on communications efficiency in FL.

\item In Section \ref{sec: communication}, communication cost reduction comes with a sacrifice in terms of either higher computation costs or lower inference accuracy. Similarly, there exist different tradeoffs to be considered in resource allocation. A scalable model is thus one that enables customization to suit varying needs. For example, the study of \cite{li2019fair} allows the FL server to calibrate levels of fairness when allocating training importance, whereas the study in \cite{yang2019energy} enables the tradeoffs between training completion time and energy expense to be calibrated by the FL system adminstrator.

\item In synchronous FL, the FL system is susceptible to the straggler effect. As such, asynchronous FL has been proposed as a solution in \cite{sprague2018asynchronous} and \cite{xie2019asynchronous}. In addition, asynchronous FL also allows participants to join the FL training halfway even while a training round is in progress. This is more reflective of practical FL settings and can be an important contributing factor towards ensuring the scalability of FL. However, synchronous FL remains to be the most common approach used due to convergence guarantees \cite{bonawitz2019towards}. Given the many advantages of asynchronous FL, new asynchronous algorithms should be investigated. In particular, for future proposed algorithms, the convergence guarantee in a non-IID setting for non-convex loss functions needs to be considered.

\item The study of incentive mechanism design is a particularly important aspect of FL. In particular, due to data privacy concerns, the FL servers are unable to check for training data quality. With the use of self-revealing mechanisms in contract theory, or through modeling the interactions between FL server and participants with game theoretic concepts, high quality data can be motivated as contributions from FL participants. However, existing studies in \cite{kangincentive2}, \cite{feng2018joint}, and \cite{kang2019incentive} generally assume that a
federation enjoys a monopoly. In particular, each system model is assumed to only consist of multiple individual participants collaborating with a sole FL server. There can be exceptions to this setting as follows: (i) the participants may be competing data owners who are reluctant to share their model parameters since the competitors
also benefit from a trained global model and (ii) the FL servers
may compete with other FL servers, i.e., model owners.
In this case, the formulation of the incentive mechanism design
will be vastly different from that proposed. A relatively novel approach has been to model the regret \cite{yuhanregret} of each FL participants in joining the various competing federations for model training. For future works, a system model with multiple competing federations can be considered together with Stackelberg games and contract theoretic approaches.

\item In this section, we have assumed that FL assures the privacy and security of participants. However, as discussed in the following section, this assumption may not hold in the presence of malicious participants or FL server. 

\end{itemize}

%======================================
%====================================== 

%====================================================================
%====================================================================

%\cite{bhagoji2018analyzing} \cite{preuveneers2018chained} \cite{nguyen2018d} \cite{abeshu2018deep} \cite{hitaj2017deep} \cite{weng2018deepchain} \cite{jiangdifferentially} \cite{geyer2017differentially} \cite{melis2019exploiting}
%\cite{kumar2017federated} \cite{yu2018federated} \cite{yang2019federated} \cite{Zhuo2019FederatedRL} \cite{bagdasaryan2018backdoor} \cite{fung2018mitigating} \cite{de2017neuron} \cite{jiang2019lightweight} \cite{kim2018device}
%\cite{chi2018privacy} \cite{zhou2018real} \cite{liu2018secure}

%==========================================================
\section{ Privacy and Security Issues}
\label{sec:security}
One of the main objectives of FL is to protect the privacy of participants, i.e., the participants only need to share parameters of the trained model instead of sharing their actual data. However, some recent research works have shown that privacy and security concerns may arise when the FL participants or FL servers are malicious in nature. In particular, this defeats the purpose of FL since the resulting global model can be corrupted, or the participants may even have their privacy compromised during model training. In this section, we discuss the following issues:

\begin{itemize}

	\item \textit{Privacy}: Even though FL does not require the exchange of data for collaborative model training, a malicious participant can still infer sensitive information, e.g., gender, occupation, and location, from other participants based on their shared models. For example, in~\cite{melis2019exploiting}, when training a binary gender classifier on the FaceScrub~\cite{Ng2014Adata} dataset, the authors show that they can infer if a certain participant's inputs are included in the dataset just from inspecting the shared model, with a very high accuracy of up to 90\%. Thus, in this section, we discuss privacy issues related to the shared models in FL and review solutions proposed to preserve the privacy of participants. 
	
	\item \textit{Security}: In FL, the participants locally train the model and share trained parameters with other participants in order to improve the accuracy of prediction. However, this process is susceptible to a variety of attacks, e.g., data and model poisoning, in which a malicious participant can send incorrect parameters or corrupted models to falsify the learning process during global aggregation. Consequently, the global model will be updated incorrectly, and the whole learning system becomes corrupted. This section discusses more details on emerging attacks in FL as well as some recent countermeasures to deal with such attacks.

\end{itemize}

%==========================================================
%==========================================================

\subsection{Privacy Issues}

%======================================
%======================================
\subsubsection{Information exploiting attacks in machine learning - A brief overview}

One of the first research works that shows the possibility of extracting information from a trained model is~\cite{Ateniese2015Hacking}. In this paper, the authors show that during the training phase, the correlations implied in the training samples are gathered inside the trained model. Thus, if the trained model is released, it can lead to an unexpected information leakage to attackers. For example, an adversary can infer the ethnicity or gender of a user from its trained voice recognition system. In~\cite{Fredrikson2015Model}, the authors develop a model-inversion algorithm which is very effective in exploiting information from decision tree-based or face recognition trained models. The idea of this approach is to compare the target feature vector with each of the possible value and then derive a weighted probability estimation which is the correct value. The experiment results reveal that by using this technique, the adversary can reconstruct an image of the victim's face from its label with a very high accuracy.

Recently, the authors in~\cite{Tramer2016Stealing} show that it is even possible for an adversary to infer information of a victim through queries to the prediction model. In particular, this occurs when a malicious participant has the access to make prediction queries on a trained model. Then, the malicious participant can use the prediction queries to extract the trained model from the data owner. More importantly, the authors point out that this kind of attack can successfully extract model information from a wide range of training models such as decision trees, logistic regressions, SVMs, and even complex training models including DNNs. Some recent research works have also demonstrated the vulnerabilities of DNN-based training models against model extraction attacks~\cite{Shokri2017Membership,Papernot2016Transferability,Laskov2014Practical}. Therefore, this raises a serious privacy concern for participants in sharing training models in FL.

%======================================
%======================================
\subsubsection{Differential privacy-based protection solutions for FL participants}

In order to protect the privacy of parameters trained by DNNs, the authors in~\cite{Abadi2016Deep} introduce a technique, called \emph{differentially private stochastic gradient descent}, which can be effectively implemented on DL algorithms. The key idea of this technique is to add some ``noise'' to the trained parameters by using a differential privacy-preserving randomized mechanism~\cite{Dwork2006Calibrating}, e.g., a Gaussian mechanism, before sending such parameters to the server. In particular, at the gradient averaging step of a normal FL participant, a Gaussian distribution is used to approximate the differentially private stochastic gradient descent. Then, during the training phase, the participant keeps calculating the probability that malicious participants can exploit information from its shared parameters. Once a predefined threshold is reached, the participant will stop its training process. In this way, the participant can mitigate the risk of revealing private information from its shared parameters. 

Inspired by this idea, the authors in~\cite{geyer2017differentially} develop an approach which can achieve a better privacy-protection solution for participants. In this approach, the authors propose two main steps to process data before sending trained parameters to the server. In particular, for each learning round, the aggregate server first selects a random number of participants to train the global model. Then, if a participant is selected to train the global model in a learning round, the participant will adopt the method proposed in~\cite{Abadi2016Deep}, i.e., using a Gaussian distribution to add noise to the trained model before sending the trained parameters to the server. In this way, a malicious participant cannot infer information of other participants by using the parameters of shared global model as it has no information regarding who has participated in the training process in each learning round.

%======================================
%======================================
\subsubsection{Collaborative training solutions}

\begin{figure}[!]
	\centering
	\includegraphics[width=\linewidth]{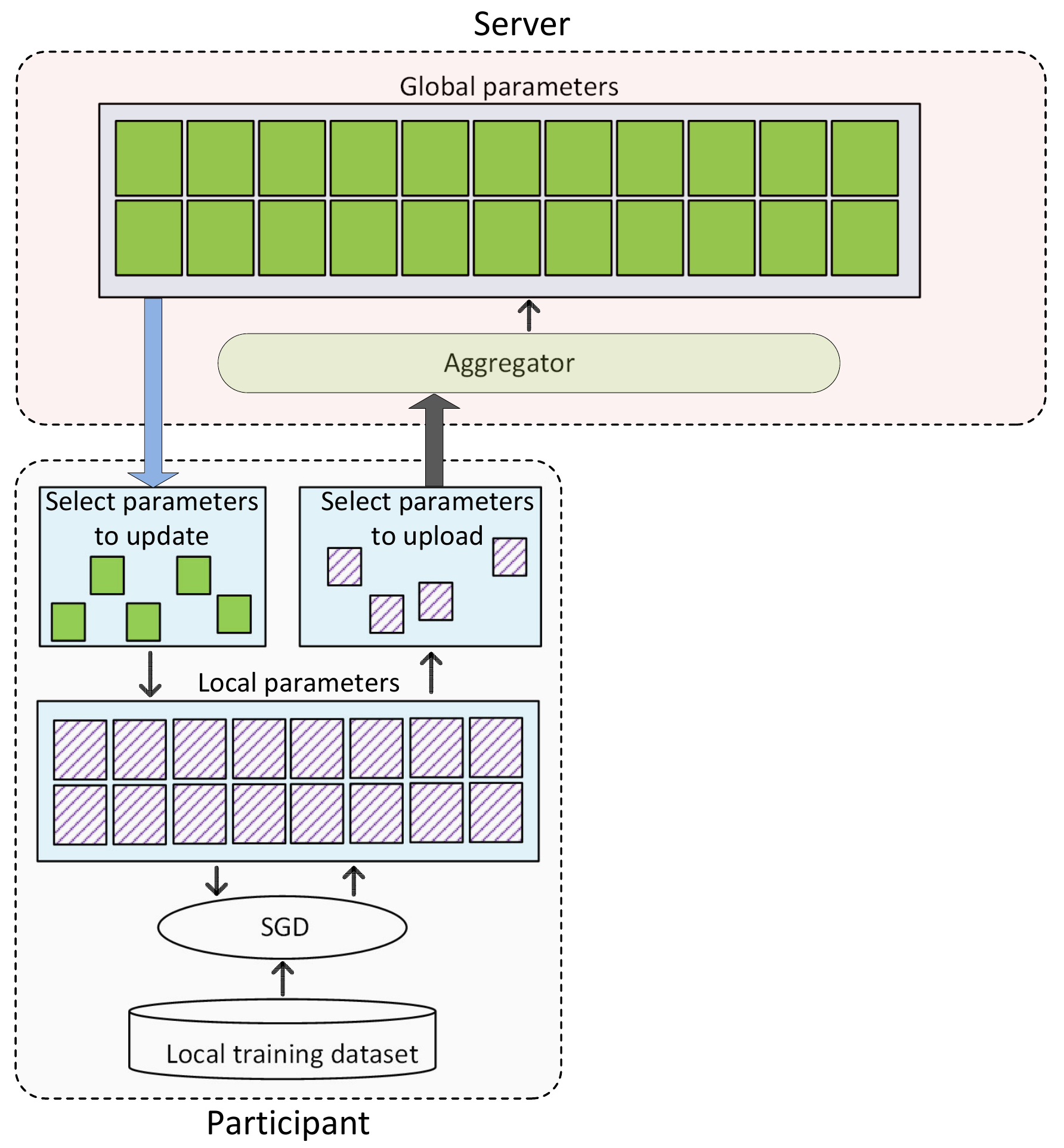}
	\caption{\small Selective parameters sharing model.}
	\label{fig:CollTrain}
\end{figure}

While DP solutions can protect private information of a honest participant from other malicious participants in FL, they only work well if the server is trustful. If the server is malicious, it can result in a more serious privacy threat to all participants in the network. Thus, the authors in~\cite{Shokri2015Privacy} introduce a collaborative DL framework to render multiple participants to learn the global model without uploading their explicit training models to the server. The key idea of this technique is that instead of uploading the whole set of trained parameters to the server and updating the whole global parameters to its local model, each participant wisely selects the number of gradients to upload and the number of parameters from the global model to update as illustrated in Fig.~\ref{fig:CollTrain}. In this way, malicious participants cannot infer explicit information from the shared model. One interesting result of this paper is that even when the participants do not share all trained parameters and do not update all parameters from the shared model, the accuracy of proposed solution is still close to that of the case when the server has all dataset to train the global model. For example, for the MNIST dataset~\cite{Lecun2001Gradient}, the accuracy of prediction model when the participants agree to share 10\% and 1\% of their parameters are respectively 99.14\% and 98.71\%, compared with 99.17\% for the centralized solution when the server has full data to train. However, the approach is yet to be tested on more complex classification tasks.

\begin{figure*}[!]
	\centering
	\includegraphics[width=\linewidth]{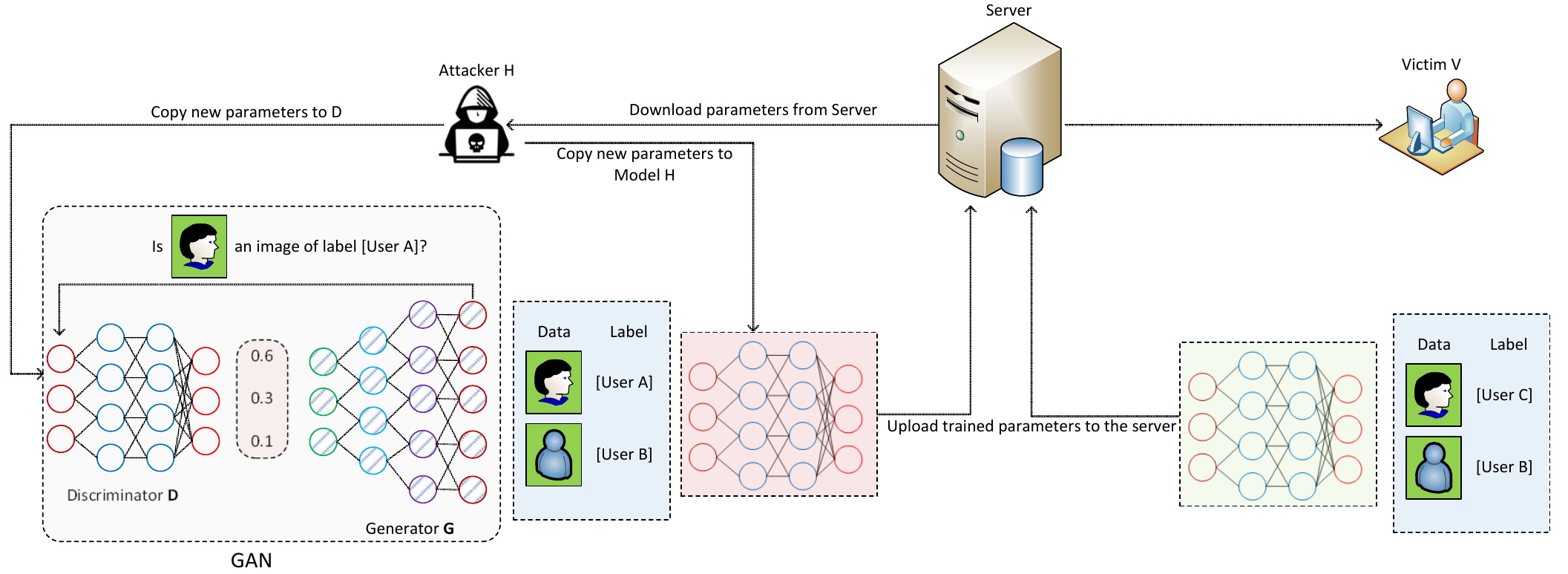}
	\caption{\small GAN Attack on collaborative deep learning.}
	\label{fig:GAN_attacks}
\end{figure*}

Although selective parameter sharing and DP solutions can make information exploiting attacks more challenging, the authors in~\cite{hitaj2017deep} show that these solutions are susceptible to a new type of attack, called powerful attack, developed based on Generative Adversarial Networks (GANs) \cite{goodfellow2014generative}. GANs is a class of ML technique which uses two neural networks, namely generator network and discriminator network, that compete with each other to train data. The generator network tries to generate the fake data by adding some ``noise'' to the real data. Then, the generated fake data is passed to the discriminator network for classification. After the training process, the GANs can generate new data with the same statistics as the training dataset. Inspired by this idea, the authors in~\cite{hitaj2017deep} develop a powerful attack which allows a malicious participant to infer sensitive information from a victim participant even with just a part of shared parameters from the victim as illustrated in Fig.~\ref{fig:GAN_attacks}. To deal with the GAN attack, the authors in~\cite{Liu2019Boosting} introduce a solution using secret sharing scheme with extreme boosting algorithm. This approach executes a lightweight secret sharing protocol before transmitting the newly trained model in plaintext to the server at each round. Thereby, other participants in the network cannot infer information from the shared model. However, the limitation of this approach is the reliance on a trusted third party to generate signature key pairs. 

Different from all aforementioned works, the authors in~\cite{Triastcyn2019Federated} introduce a collaborative training model in which all participants cooperate to train a federated GANs model. The key idea of this method is that the federated GANs model can generate artificial data that can replace participants' real data, and thus protecting the privacy of real data for the honest participants. In particular, in order to guarantee participants' data privacy while still maintaining flexibility in training tasks, this approach produces a federated generative model. This model can output artificial data that does not belong to any real user in particular, but comes from the common cross-user data distribution. As a result, this approach can significantly reduce the possibility of malicious exploitation of information from real data. However, this approach inherits existing limitations of GANs, e.g., training instability due to the generated fake data, which can dramatically reduce the performance of collaborative learning models.

\subsubsection{Encryption-based Solutions}
%======================================
%======================================

Encryption is an effective way to protect data privacy of the participants when they want to share the trained parameters in FL. In~\cite{Aono2017Privacy}, the homomorphic encryption technique is introduced to protect privacy of participants' shared parameters from a honest-but-curious server. A honest-but-curious server is defined to be a user who wants to extract information from the participants' shared parameters, but keeps all operations in FL in proper working condition. The idea of this solution is that the participants' trained parameters will be encrypted using the homomorphic encryption technique before they are sent to the server. This approach is effective in protecting sensitive information from the curious server, and also achieves the same accuracy as that of the centralized DL algorithm. A similar concept is also presented in~\cite{bonawitz2017practical} with secret sharing mechanism used to protect information of FL participants. 

Although both the encryption techniques presented in~\cite{Aono2017Privacy} and~\cite{bonawitz2017practical} can prevent the curious server from extracting information, they require multi-round communications and cannot preclude collusions between the server and participants. Thus, the authors in~\cite{Hao2019Towards} propose a hybrid solution which integrates both additively homomorphic encryption and DP in FL. In particular, before the trained parameters are sent to the server, they will be encrypted using the additively homomorphic encryption mechanism together with intentional noises to perturb the original parameters. As a result, this hybrid scheme can simultaneously prevent the curious server from exploiting information as well as solve the collusion problem between the server and malicious participants. However, in this paper, the authors do not compare the accuracy of the proposed approach with the case without homomorphic encryption and DP. Thus, the performance of proposed approach, i.e., in terms of model accuracy, is not clear.

%==========================================================
%==========================================================
\subsection{Security Issues}
\label{sec:securityissues}

%======================================
%======================================
\subsubsection{Data Poisoning Attacks}

In FL, a participant trains its data and sends the trained model to the server for further processing. In this case, it is intractable for the server to check the real training data of a participant. Thus, a malicious participant can poison the global model by creating \emph{dirty-label} data to train the global model with the aim of generating falsified parameters. For example, a malicious participant can generate a number of samples, e.g., photos, under a designed label, e.g., a clothing branch, and use them to train the global model to achieve its business goal, e.g., the prediction model shows results of the targeted clothing branch. Dirty-label data poisoning attacks are demonstrated to achieve high misclassifications in DL processes, up to 90\%, when a malicious participant injects relatively few dirty-label samples (around 50) to the training dataset~\cite{Chen2017Targeted}. This calls for urgent solutions to deal with data poisoning attacks in FL.

In~\cite{fung2018mitigating}, the authors investigate impacts of a sybil-based data poisoning attack to a FL system. In particular, for the sybil attack, a malicious participant tries to improve the effectiveness of data poisoning in training the global model by creating multiple malicious participants. In Table~\ref{tab:Syblil}, the authors show that with only two malicious participants, the attack success rate can achieve up to 96.2\%, and now the FL model is unable to correctly classify the image of ``1'' (instead it always incorrectly predicts them to be the image of ``7''). To mitigate sybil attacks, the authors then propose a defense strategy, namely FoolsGold. The key idea of this approach is that honest participants can be distinguished from sybil participants based on their updated gradients. Specifically, in the non-IID FL setting, each participant's training data has its own particularities, and sybil participants will contribute gradients that appear more similar to each other than those of other honest participants. With FoolsGold, the system can defend the sybil data poisoning attack with minimal changes to the conventional FL process and without requiring any auxiliary information outside of the learning process. Through simulations results on 3 diverse datasets (MNIST~\cite{Lecun2001Gradient}, KDDCup~\cite{Dheeru2017UCI}, Amazon Reviews~\cite{Dheeru2017UCI}), the authors show that FoolsGold can mitigate the attack under a variety of conditions, including different distributions of participant data, varying poisoning targets, and various attack strategies.

\begin{table}[!]
	\centering
	\caption{\small The accuracy and attack success rates for no-attack scenario and attacks with 1 and 2 sybils in a FL system with MNIST dataset~\cite{Lecun2001Gradient}.}
	\label{tab:Syblil}
	\begin{tabular}{||c||c|c|c||} \hline
		& \textbf{Baseline} & \textbf{Attack 1} & \textbf{Attack 2} \\ 
		\hline
		\hline 		
		Number of honest participants & 10 & 10 & 10   			\\ \cline{1-4}
		Number of sybil participants & 0 & 1 & 2     			\\ \cline{1-4}
		The accuracy (digits: 0, 2-9) & 90.2\% & 89.4\% & 88.8\%    \\ \cline{1-4}
		The accuracy (digit: 1) & 96.5\% & 60.7\% & 0.0\%    \\ \cline{1-4}
		\textbf{Attack success rate} & 0.0\% & 35.9\% & 96.2\%    \\ \cline{1-4}
		\hline\end{tabular}
\end{table}

%======================================
%======================================
\subsubsection{Model Poisoning Attacks}

Unlike data poisoning attacks which aim to generate fake data to cause adverse impacts to the global model, a model poisoning attack attempts to directly poison the global model that it sends to the server for aggregation. As shown in~\cite{bhagoji2018analyzing} and \cite{bagdasaryan2018backdoor}, model poisoning attacks are much more effective than those of data poisoning attacks, especially for large-scale FL with many participants. The reason is that for data poisoning attacks, a malicious participant's updates are scaled based on its dataset and the number of participants in the federation. However, for model poisoning attacks, a malicious participant can modify the updated model, which is sent to the server for aggregation, directly. As a result, even with one single attacker, the whole global model can be poisoned. The simulation results in~\cite{bhagoji2018analyzing} also confirm that even a highly constrained adversary with limited training data can achieve high success rate in performing model poisoning attacks. Thus, solutions to protect the global model from model poisoning attacks have to be developed. 

In~\cite{bhagoji2018analyzing}, some solutions are suggested to prevent model poisoning attacks. Firstly, based on an updated model shared from a participant, the server can check whether the shared model can help to improve the global model's performance or not. If not, the participant will be marked to be a potential attacker, and after few rounds of observing the updated model from this participant, the server can determine whether this is a malicious participant or not. The second solution is based on the comparison among the updated models shared by the participants. In particular, if an updated model from a participant is too different from the others, the participant can potentially be a malicious one. Then, the server will continue observing updates from this participant before it can determine whether this is a malicious user or not. However, model poisoning attacks are extremely difficult to prevent because when training with millions of participants, it is intractable to evaluate the improvement from every single participant. As such, more effective solutions need to be further investigated. 

In~\cite{bagdasaryan2018backdoor}, the authors introduce a more effective model poisoning attack which is demonstrated to achieve 100\% accuracy on the attacker's task within just a single learning round. In particular, a malicious participant can share its poisoned model which not only is trained for its intentional purpose, but which also contains a backdoor function. In this paper, the authors consider to use a semantic backdoor function to inject into the global model. The reason is that this function can make the global model misclassify even without a need to modify the input data of the malicious participant. For example, an image classification backdoor function can inject an attacker-chosen label to all images with some certain features, e.g., all dogs with black stripes can be misclassifed to be cats. In the simulations, the authors show that this attack can greatly outperform other conventional FL data poisoning attacks. For example, in a word-prediction task with 80,000 total participants, compromising just eight of them is enough to achieve 50\% backdoor accuracy, as compared to 400 malicious participants needed to perform the data-poisoning attack.

%======================================
%======================================
\subsubsection{Free-Riding Attacks}

\begin{figure}[!]
	\centering
	\includegraphics[width=\linewidth]{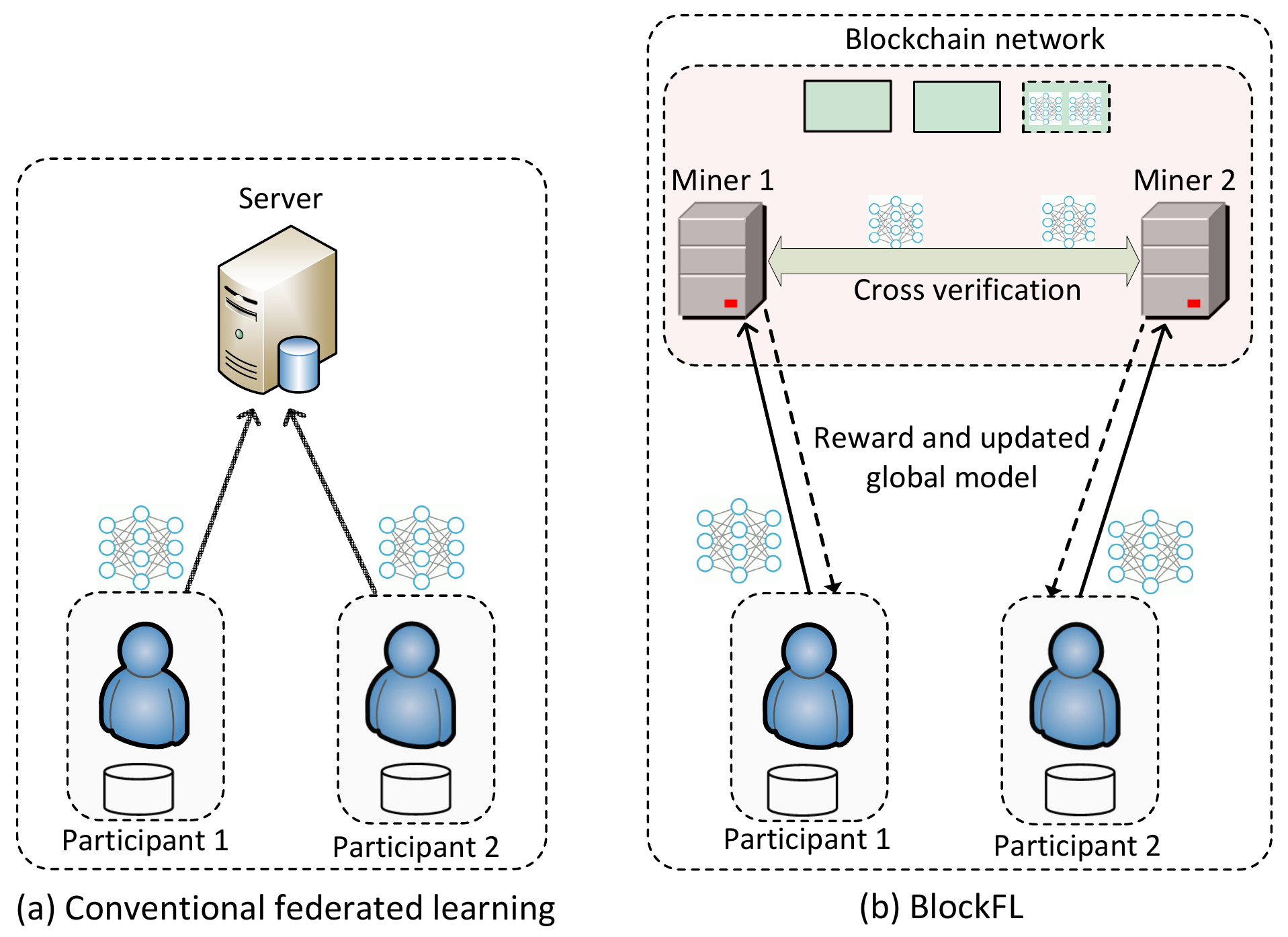}
	\caption{\small An illustration of (a) conventional FL and (b) the proposed BlockFL architectures.}
	\label{fig:BlockFL}
\end{figure}

Free-riding is another attack in FL that occurs when a participant wants to benefit from the global model without contributing to the learning process. The malicious participant, i.e., free-rider, can pretend that it has very small number of samples to train or it can select a small set of its real dataset to train, e.g., to save its resources. As a result, the honest participants need to contribute more resources in the FL training process. To address this problem, the authors in~\cite{kim2018device} introduce a blockchain-based FL architecture, called BlockFL, in which the participants' local learning model updates are exchanged and verified by leveraging the blockchain technology. In particular, each participant trains and sends the trained global model to its associated miner in the blockchain network and then receives a reward that is proportional to the number of trained data samples as illustrated in Fig.~\ref{fig:BlockFL}. In this way, this framework can not only prevent the participants from free-riding, but also incentivize all participants to contribute to the learning process. A similar blockchain-based model is also introduced in~\cite{weng2018deepchain} to provide data confidentiality, computation auditability, and incentives for the participants of FL. However, the utilization of the blockchain technology implies the incurrence of a significant cost for implementing and maintaining miners to operate the blockchain network. Furthermore, consensus protocols used in blockchain networks, e.g., proof-of-work (PoW), can cause a long delay in information exchange, and thus they may not be appropriate to implement on FL models.

\begin{table*}[!]
	\caption{\small A summary of attacks and countermeasures in FL.}
	\label{tab:Sec_Pri}
	\scriptsize	
	\begin{centering}
		\begin{tabular}{|>{\centering\arraybackslash}m{2cm}|>{\centering\arraybackslash}m{4.5cm}|>{\centering\arraybackslash}m{10.5cm}|}
			\hline
			\cellcolor{mygray} & \cellcolor{mygray} &\cellcolor{mygray} \tabularnewline
			\cellcolor{mygray} \multirow{-2 }{*}{\textbf{Attack Types}} &\cellcolor{mygray} \multirow{-2}{*} {\textbf{Attack Method}} &\cellcolor{mygray} \multirow{-2}{*} {\textbf{Countermeasures}} \tabularnewline
			\hline
			\hline
			Information exploiting attacks (privacy issues) & Attackers try to illegally exploit information from the shared model.  &  \begin{itemize}
				\item \emph{Differentially private stochastic gradient descent}:  Add ``noise'' to the trained parameters by using a differential privacy-preserving randomized mechanism~\cite{Abadi2016Deep}.
				\item \emph{Differentially private and selective participants}:  Add ``noise'' to the trained parameters and select randomly participants to train global model in each round~\cite{geyer2017differentially}.
				\item \emph{Selective parameter sharing}: Each participant wisely selects the number of gradients to upload and the number of parameters from the global model to update~\cite{Shokri2015Privacy}.
				\item \emph{Secrete sharing scheme with extreme boosting algorithm}: This approach executes a lightweight secret sharing protocol before transmitting the newly trained model in plaintext to the server at each round~\cite{Liu2019Boosting}.
				\item \emph{GAN model training}: All participants are cooperative to train a federated GANs model~\cite{Triastcyn2019Federated}.
			\end{itemize} \\
			\hline
			Data poisoning attacks & Attackers poison the global model by creating \emph{dirty-label} data and use such data to train the global model.  &  \begin{itemize} \item \emph{FoolsGoal}: Distinguish honest participants based on their updated gradients. It is based on the fact that in the non-IID FL setting, each participant's training data has its own particularities, and malicious participants will contribute gradients that appear more similar to each other than those of the honest participants~\cite{fung2018mitigating}. \end{itemize} \\
			\hline
			Model poisoning attacks & Attackers attempt to directly poison the global model that they send to the server for aggregation.  &  \begin{itemize}
				\item Based on an updated model shared from a participant, the server can check whether the shared model can help to improve the global model's performance or not. If not, the participant will be marked to be a potential attacker~\cite{bhagoji2018analyzing}.
				\item Compare among the updated global models shared by the participants, and if an updated global model from a participant is too different from others, it could be a potential malicious participant~\cite{bhagoji2018analyzing}.
			\end{itemize}  \\
			\hline
			Free-riding attacks & Attackers benefit from the global model without contributing to the learning process, e.g., by pretending that they have very small number of samples to train.  &  \begin{itemize}
				\item \emph{BlockFL}: Participants' local learning model updates are exchanged and verified by leveraging blockchain technology. In particular, each participant trains and sends the trained global model to its associated miner in the blockchain network and then receives a reward that is proportional to the number of trained data samples~\cite{kim2018device}.
			\end{itemize}  \\
			\hline			
		\end{tabular}
		\par\end{centering}
\end{table*}

%====================================================================
%====================================================================

\subsection{Summary and Lessons Learned}

In this section, we have discussed two key issues, i.e., privacy and security, when trained models are exchanged in FL. In general, it is believed that FL is an effective privacy-preserving learning solution for participants to perform collaborative model training. However, in this section, we have shown that a malicious participant can exploit the process and gain access to sensitive information of other participants. Furthermore, we have also shown that by using the shared model in FL, an attacker can perform attacks which can not only breakdown the whole learning system, but also falsify the trained model to achieve its malicious goal. In addition, solutions to deal with these issues have also been reviewed, which are especially important in order to guide FL system administrators in designing and implementing the appropriate countermeasures. We summarize the key information of attacks and their corresponding countermeasures in Table VII.

%====================================================================
%====================================================================

\section{Applications of Federated Learning for Mobile Edge Computing}
\label{sec:application}

In the aforementioned studies, we have discussed the issues pertaining to the implementation of FL as an enabling technology that allows collaborative learning at mobile edge networks. In this section, we focus instead on the applications of FL for mobile edge network optimization. 

As highlighted by the authors in \cite{niknam2019federated}, the increasing complexity and heterogeneity of wireless networks enhance the appeal of adopting a data-driven ML based approach \cite{simeone2018very} for optimizing system designs and resource allocation decision making for mobile edge networks. However, as discussed in previous sections, the private data of users may be sensitive in nature. As such, existing learning based approach can be combined with FL for privacy-preserving applications.

In this section, we consider four applications of FL in edge computing:

\begin{itemize}

\item \textit{Cyberattack Detection:} The ubiquity of IoT devices and increasing sophistication of cyberattacks \cite{stojmenovic2016overview} imply that there is a need to improve existing cyberattack detection tools. Recently, DL has been widely successful in cyberattack detection. Coupled with FL, cyberattack detection models can be learned collaboratively while maintaining user privacy.

\item \textit{Edge Caching and Computation Offloading:} Given the computation and storage capacity constraints of edge servers, some computationally intensive tasks of end devices have to be offloaded to the remote cloud server for computation. In addition, commonly requested files or services should be placed on edge servers for faster retrieval, i.e., users do not have to communicate with the remote cloud when they want to access these files or services. As such, an optimal caching and computation offloading scheme can be collaboratively learned and optimized with FL.

\item \textit{Base Station Association:} In a dense network, it is important to optimize base station association so as to limit interference faced by users. However, traditional learning based approaches that utilize user data often assume that such data is centrally available. Given user privacy constraints, an FL based approach can be adopted instead.

\item \textit{Vehicular Networks:} The Internet of Vehicles (IoV) \cite{gerla2014internet} features smart vehicles with data collection, computation and communication capabilities for relevant functions, e.g., navigation and traffic management. However, this wealth of knowledge is again, private and sensitive in nature since it can reveal the driver's location and personal information. In this section, we discuss the use of an FL based approach in traffic queue length prediction and energy demand in electric vehicle charging stations done at the edge of IoV networks.

\end{itemize}

% \usepackage{multirow}
\begin{table*}[]
\centering \caption{\small FL based approaches for mobile edge network optimization. \label{tab:application}}
\begin{tabular}{|l|l|l|}
\hline
\textbf{Applications}                                             & \textbf{Ref.} & \textbf{Description}                                             \\ \hline
\multirow{3}{*}{Cyberattack Detection}                   & \cite{abeshu2018deep}      & Cyberattack detection with edge nodes as participants   \\ \cline{2-3} 
                                                         & \cite{nguyen2018d}      & Cyberattack detection with IoT gateways as participants \\ \cline{2-3} 
                                                         & \cite{preuveneers2018chained}     & Blockchain to store model updates                       \\ \hline
\multirow{4}{*}{Edge caching and computation offloading} & \cite{wang2018edge}     & DRL for caching and offloading in UEs                   \\ \cline{2-3} 
                                                         & \cite{ren2019federated}     & DRL for computation offloading in IoT devices           \\ \cline{2-3} 
                                                         & \cite{yu2018federated}     & Stacked autoencoder learning for proactive caching      \\ \cline{2-3} 
                                                         & \cite{qian2019privacy}     & Greedy algorithm to optimize service placement schemes  \\ \hline
\multirow{2}{*}{Base station assoication}                & \cite{Chen2018}      & Deep echo state networks for VR application             \\ \cline{2-3} 
                                                         & \cite{hamidouche2018collaborative}      & Mean field game with imitation for cell association     \\ \hline
\multirow{2}{*}{Vehicular networks}                      & \cite{samarakoon2018federated}     & Extreme value theory for large queue length prediction  \\ \cline{2-3} 
                                                         &  \cite{saputra2019energy}      & Energy demand learning in electric vehicular networks   \\ \hline
\end{tabular}
\end{table*}

\subsection{Cyberattack Detection}

Cyberattack detection is one of the most important steps to promptly prevent and mitigate serious consequences of attacks in mobile edge networks. Among different approaches to detect cyberattacks, DL is considered to be the most effective tool to detect a wide range of attacks with high accuracy. In~\cite{nguyen2018cyberattack}, the authors show that DL can outperform all conventional ML techniques with very high accuracy in detecting intrusions on three datasets, i.e., KDDcup 1999, NSL-KDD \cite{nsl}, and UNSW-NB15 \cite{moustafa2015unsw}. However, the detection accuracy of solutions based on DL depends very much on the available datasets. Specifically, DL algorithm only can outperform other ML techniques when given sufficient data to train. However, this data may be sensitive in nature. Therefore, some FL-based attack detection models for mobile edge networks have been introduced recently to address this problem.

\begin{figure}[!]
	\centering
	\includegraphics[width=\columnwidth]{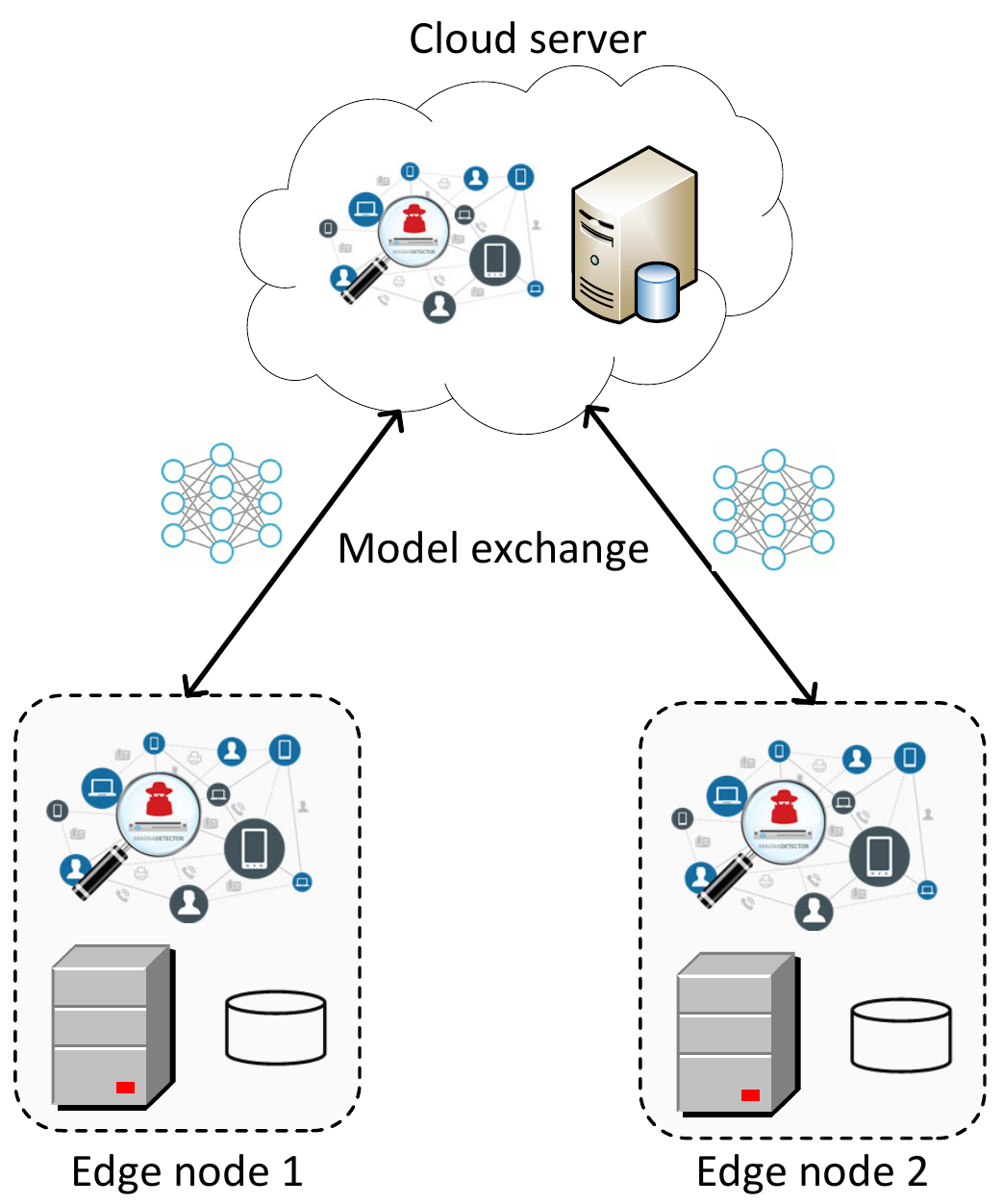}
	\caption{\small FL-based attack detection architecture for IoT edge networks.}
	\label{fig:IDS_FL}
\end{figure}

In~\cite{abeshu2018deep}, the authors propose a cyberattack detection model for an edge network empowered by FL. In this model, each edge node operates as a participant who owns a set of data for intrusion detection. To improve the accuracy in detecting attacks, after training the global model, each participant will send its trained model to the FL server. The server will aggregate all parameters from the participants and send the updated global model back to all the participants as illustrated in Fig.~\ref{fig:IDS_FL}. In this way, each edge node can learn from other edge nodes without a need of sharing its real data. As a result, this method can not only improve accuracy in detecting attacks, but also enhance the privacy of intrusion data at the edge nodes and reduce traffic load for the whole network. A similar idea is also presented in~\cite{nguyen2018d} in which IoT gateways operate as FL participants and an IoT security service provider works as a server node to aggregate trained models shared by the participants. The authors in~\cite{nguyen2018d} show empirically that by using FL, the system can successfully detect 95.6\% of attacks in approximately 257 ms without raising any false alarm when evaluated in a real-world smart home deployment setting.

In both~\cite{abeshu2018deep} and~\cite{nguyen2018d}, it is assumed that the participants, i.e., edge nodes and IoT gateways, are honest, and they are willing to contribute in training their updated model parameters. However, if some of the participants are malicious, they can make the whole intrusion detection corrupted. Thus, the authors in~\cite{preuveneers2018chained} propose to use blockchain technology in managing data shared by the participants. By using the blockchain, all incremental updates to the anomaly detection ML model are stored in the ledger, and thus a malicious participant can be easily identified. Furthermore, based on shared models from honest participants stored in the ledger, the intrusion detection system can easily recover the proper global model if the current global model is poisoned.

\subsection{Edge Caching and Computation Offloading}

\begin{figure}[!]
	\centering
	\includegraphics[width=\linewidth]{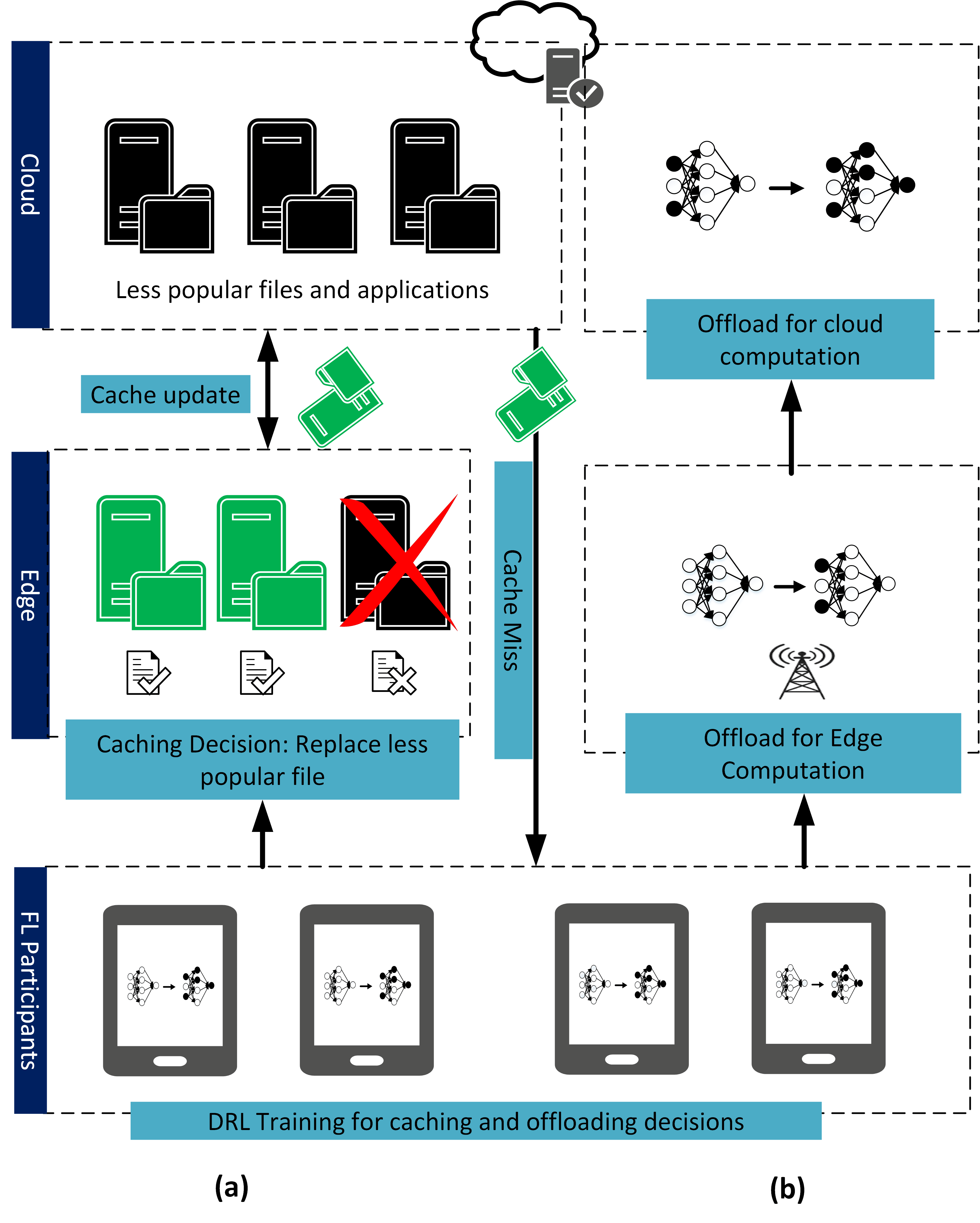}
	\caption{\small FL-based (a) caching and (b) computation offloading.}
	\label{fig:caching}
\end{figure}

To account for the dynamic and time-varying conditions in a MEC system, the authors in \cite{wang2018edge} propose the use of DRL with FL to optimize caching and computation offloading decisions in an MEC system. The MEC system consists of a set of user equipments (UEs) covered by base stations. For caching, the DRL agent makes the decision to cache or not to cache the downloaded file, and which local file to replace should caching occur. For computation offloading, the UEs can choose to either offload
computation tasks to the edge node via wireless channels,
or perform the tasks locally. This caching and offloading decision process is illustrated in Fig. \ref{fig:caching}. The states of the MEC system include wireless network conditions,
UE energy consumption, and task queuing states, whereas the reward function is
defined as quality of experience (QoE) of the UEs. Given the large state and action space in the MEC environment, a DDQN approach is adopted. To protect the privacy of users, an FL approach is proposed in which training can occur with data remaining on the UEs. In addition, existing FL algorithms, e.g., \textit{FedAvg} \cite{mcmahan2016communication}, can also ensure that training is robust to the unbalanced and non-IID data of the UEs. The simulation
results show that the DDQN with FL approach achieves similar average
utilities among UEs as compared to the centralized DDQN
approach, while consuming less communication resources and
preserving user privacy. However, the simulations are only performed with $10$ UEs. If the implementation is expanded to target a larger number of heterogeneous UEs, there can be significant delays in the training process especially since the training of a DRL model is computationally intensive. As an extension, transfer learning \cite{pan2009survey} can be used to increase the efficiency of training, i.e., training is not initialized from scratch.

Similar to \cite{wang2018edge}, the authors in \cite{ren2019federated} propose the use of DRL in optimizing computation offloading decisions in IoT systems. The system model consists of IoT devices and edge nodes. The IoT devices can harvest energy units \cite{priya2009energy} from the edge nodes to be stored in the energy queue. In addition, an IoT device also maintains a local task queue with unprocessed and unsuccessfully processed tasks. These tasks can be processed locally or offloaded to the edge nodes for processing, in a First In First Out (FIFO) order \cite{li2018learning}. In the DRL problem formulation, the network states are defined to be a function of energy queue length, task execution delay, task handover delay from edge node association, and channel gain between the IoT device and edge nodes. A task can fail to be executed, e.g., when there is insufficient energy units or communication bandwidth for computation offloading. The utility considered is a function of task execution delay, task queuing delay, number of failed tasks and penalty of execution failure. The DRL agent makes the decision to either offload computation to the edge nodes or perform computation locally. To ensure privacy of users, the agent is trained without users having to upload their own data to a centralized server. In each training round, a random set of IoT devices are selected to download the model parameters of the DRL agent from the edge networks. The model parameters are then updated using their own data, e.g., energy resource level, channel gain, and local sensing data. Then, the updated parameters of the DRL agent are sent to the edge nodes for model aggregation. The simulation results show that the FL based approach can achieve same levels of total utility as the centralized DRL approach. This is robust to varying task generation probabilities. In addition, when task generation probabilities are higher, i.e., there are more tasks for computation in the IoT device, the FL based scheme can achieve a lower number of dropped tasks and shorter queuing delay than the centralized DRL scheme. However, the simulation only involves $15$ IoT devices serviced by relatively many edge nodes. To better reflect practical scenarios where fewer edge nodes have to cover several IoT devices, further studies can be conducted on optimizing the edge-IoT collaboration. For example, the limited communication bandwidth can cause significant task handover delay during computation offloading. In addition, with more IoT devices, the DRL training will take a longer time to converge especially since the devices have heterogeneous computation capabilities.

Instead of using a DRL approach, the authors in \cite{yu2018federated} propose the use of an FL based stacked autoencoder learning model, i.e., FL based proactive content caching scheme (FPCC), to predict content popularity for optimized caching while protecting user privacy. In the system model, each user is equipped with a mobile device that connects to the base station that covers its geographical location. Using a stacked autoencoder learning model, the latent representation of a user's information, e.g., location, and file rating, i.e., content request history, is learned. Then, a similarity matrix between the user and its historically requested files is obtained in which each element of the matrix represents the distance between the user and the file. Based on this similarity matrix, the $K$ nearest neighbours of each user are determined, and the similarity between the user's historical watch list and the neighbours' are computed. An aggregation approach is then used to predict the most popular files for caching, i.e., files with highest similarity scores across all users. Being the most popular files across users that are most frequently retrieved, the cached files need not be re-downloaded from its source server everytime it is demanded. To protect the privacy of users, FL is adopted to learn the parameters of the stacked autoencoder without the user having to reveal its personal information or its content request history to the FL server. In each training round, the user first downloads a global model from the FL server. Then, the model is trained and updated using their local data. The updated models are subsequently uploaded to the FL server and aggregated using the \textit{FedAvg} algorithm. The simulation results show that the proposed FPCC scheme could achieve the highest cache efficiency, i.e, the ratio of cached files matching user requests, as compared to other caching methods such as the Thompson sampling methods \cite{chapelle2011empirical}. In addition, privacy of the user is preserved.

The authors in \cite{qian2019privacy} introduce a privacy-aware service placement scheme to deploy user-preferred services on edge servers with consideration for resource constraints in the edge cloud. The system model consists of a mobile edge cloud serving various mobile devices. The user's preference model is first built based on information such as number of times of requests for a service, and other user context information, e.g., ages and locations. However, since this can involve sensitive personal information, an FL based approach is proposed to train the preference model while keeping users' data on their personal devices.  Then, an optimization problem is formulated in which the objective is to maximize quantity of services demanded from the edge based on user preferences, subject to constraints of storage capacity, computation capability, uplink and downloading bandwidth. The optimization problem is then solved using a greedy algorithm, i.e., the service which most improves the objective function is added till resource constraints are met. The simulation results show that the proposed scheme can outperform the popular service placement scheme, i.e., where only the most popular services are placed on the edge cloud, in terms of number of requests processed on edge clouds since it also considers the aforementioned resource constraints in maximizing quantity of services.

\subsection{Base Station Association}

The authors in \cite{Chen2018} propose an FL based deep echo state networks (ESNs) approach to minimize breaks in presence (BIPs) \cite{chung2010analysis} for users of virtual reality (VR) applications. A BIP event can be a result of delay in information transmission which can be caused when the user's body movements obstruct the wireless link. BIPs cause the user to be aware that they are in a virtual environment, thus reducing their quality of experience. As such, a user association policy has to be designed such that BIPs are minimized. The system model consists of base stations that cover a set of VR users. The base stations receive uploaded tracking information from each associated user, e.g., physical location and orientation, while the users download VR videos for their use in the VR application. For data transmission, the VR users have to associate with one of the base stations. As such, a minimization problem is formulated where BIPs are minimized with respect to expected locations and orientations of the VR user. To derive a prediction of user locations and orientations, the base station has to rely on the historical information of users. However, the historical information stored at each base station only collects partial data from each user, i.e., a user connects to multiple base stations and its data is distributed across them. As such, an FL based approach is implemented whereby each base station first trains a local model using its partial data. Then, the local models are aggregated to form a global model capable of generalization, i.e., comprehensively predicting a user's mobility and orientations. The simulation results show that the federated ESN algorithm can achieve lower BIPs experienced by users as compared to the centralized ESN algorithm proposed in \cite{chen2017caching}, since a centralized approach only makes partial prediction with the incomplete data from sole base stations, whereas the federated ESN approach can make predictions based on a model learned collaboratively from more complete data.

Following the ubiquity of IoT devices, the traditional cloud-based approach may no longer be sufficient to cater to dense cellular networks. As computation and storage moves to the edge of networks, the association of users to base stations are increasingly important to facilitate efficient ML model training among the end users. To this end, the authors in \cite{hamidouche2018collaborative} consider solving the problem of cell association in dense wireless networks with a collaborative learning approach. 
In the system model, the base stations cover a set of users in an LTE cellular system. In a cellular system, users are likely to face similar channel conditions as their neighbors and thus can benefit from learning from their neighbours that are already associated with base stations. As such, the cell association problem is formulated as a mean-field game (MFG) with imitation \cite{gueant2011mean} in which each user maximizes its own throughput while minimizing the cost of imitation.
The MFG is further
reduced into a single-user Markov decision process that is then solved
by a neural Q-learning algorithm. In
most other proposed solution for cell association, it is assumed that
all information is known to the base stations and users. However, given
privacy concerns, the assumption of information sharing may
not be practical. As such, a collaborative learning approach can be considered where only the outcome of the learning algorithm is exchanged
during the learning process whereas usage data
is kept locally in each user's own device. The simulation results show that imitating users can attain higher utility within a shorter
training duration as compared to non-imitating users.

\subsection{Vehicular Networks}

Ultra reliable low latency communication (URLLC) in vehicular networks
is an essential prerequisite towards developing an intelligent transport system.
However, existing radio resource management techniques do
not account for rare events such as large queue lengths at
the tail-end distribution. To model the occurrence of such low
probability events, the authors in \cite{samarakoon2018federated} propose the use of extreme value theory (EVT) \cite{de2007extreme}. The approach requires
sufficient samples of queue state information (QSI) and data
exchange among vehicles. As such, an FL approach is proposed in which
vehicular users (VUEs) train the learning model with data kept locally and
upload only their updated model parameters to the roadside units (RSU). The
RSU then averages out the model parameters and return an updated
global model to the VUEs. In a synchronous approach, all
VUEs upload their models at the end of a prespecified interval. However, the simultaneous uploading by multiple vehicles can lead to delays in communication.
In contrast for an asynchronous approach, each VUE only
evaluates and uploads their model parameters after a predefined number
of QSI samples are collected. The global model is also updated whenever a local update is received, thus reducing communication delays. To further reduce overhead,
Lyapunov optimization \cite{neely2010stochastic} for power allocation is also utilized. The
simulation results show that under this framework, there is
a reduction of the number of vehicles experiencing large queue lengths
whereas FL can ensure minimal data exchange relative to a
centralized approach.

Apart from QSI, the vehicles in vehicular networks are also exposed to a wealth of useful captured images that can be adopted to build better inference models, e.g., for traffic optimization. However, these images are sensitive in nature since they can give away the location information of vehicular clients. As such, an FL approach can be used to facilitate collaborative ML while ensuring privacy preservation. However, the images captured by vehicles are often varying in quality due to motion blurs. In addition, another source of heterogeneity is the difference in computing capabilities of vehicles. Given the information asymmetry involved, the authors in \cite{ye2020federated} propose a multi-dimensional contract design in which the FL server designs contract bundles comprising varying levels of data quality, compute resources, and contractual payoffs. Then, the vehicular client chooses the contract bundle that maximizes its utility, in accordance to its hidden type. Similar to the results in \cite{kang2019incentive}, the simulation results show that the FL server derives greatest utility under the proposed contract theoretic approach, in contrast to the linear pricing or Stackelberg game approach.

The authors in \cite{saputra2019energy} propose a federated energy demand learning (FEDL) approach to manage energy resources in charging stations (CSs) for electric vehicles (EVs). When a large number of EVs congregate at a CS, this can lead to energy transfer congestion. To resolve this, energy is supplied from the power grids and reserved in advance to meet the real-time demands from the EVs \cite{you2014efficient}, rather than having the CSs request for energy from the power grid only upon receiving charging requests. As such, there is a need to forecast energy demand for EV networks using historical charging data. However, this data is usually stored separately at each of the CS that the EVs utilize and is private in nature. As such, an FEDL approach is adopted in which each CS trains the demand prediction model on its own dataset before sending only the gradient information to the charging station provider (CSP). Then, the gradient information from the CS is aggregated for global model training. To further improve model accuracy, the CSs are clustered using the constrained K-means algorithm \cite{bradley2000constrained} based on their physical locations. The clustering-based FEDL reduces the cost of biased prediction \cite{li2018implemented}. The simulation results show that the root mean squared error of a clustered FEDL model is lower than conventional ML algorithms, e.g., multi-layer perceptron regressor \cite{boutaba2018comprehensive}. However, the privacy of user data is still not protected by this approach, since user data is stored in each of the CS. As an extension, the user data can possibly be stored in each EVs separately, and model training can be conducted in the EVs rather than the CSs. This can allow more user features to be considered to enhance the accuracy of EDL, e.g., user consumption habits.

\textit{Summary:} In this section, we discuss that FL can also be used for mobile edge network optimization. In particular, DL and DRL approaches are suitable for modelling the dynamic environment of increasingly complex edge networks but require sufficient data for training. With FL, model training can be carried out while preserving the privacy of users. A summary of the approaches are presented in Table \ref{tab:application}.

\section{Challenges and Future Research Directions}
\label{sec:challenges_open_issues}
%====================================================================
%====================================================================
Apart from the aforementioned issues, there are still challenges new research directions in deploying FL at scale to be discussed as follows.

\begin{itemize}

\item \textit{Dropped participants:} The approaches discussed in Section \ref{sec:resource}, e.g., \cite{nishio2018client}, \cite{yoshida2019hybrid}, and \cite{anh2018efficient}, propose new algorithms for participant selection and resource allocation to address the training bottleneck and resource heterogeneity. In these approaches, the wireless connections of participants are assumed to be always available. However, in practice, participating mobile devices may go offline and can drop out from the FL system due to connectivity or energy constraints. A large number of dropped devices from the training participation can significantly degrade the performance \cite{mcmahan2016communication}, e.g., accuracy and convergence speed, of the FL system. New FL algorithms need to be robust to device drop out in the networks and anticipate the scenarios in which only a small number of participants are left connected to participate in a training round. One potential solution is that the FL model owner provides free dedicated/special connection, e.g., cellular connections, as an incentive to the participants to avoid drop out. 

\item \textit{Privacy concerns:} FL is able to protect the privacy of each participants since the model training may be conducted locally, with just the model parameters exchanged with the FL server. However, as specified in \cite{Ateniese2015Hacking},~\cite{Fredrikson2015Model}, and \cite{Tramer2016Stealing}, communicating the model updates during the training process can still reveal sensitive information to an adversary or a third-party. The current approaches propose security solutions such as DP, e.g., \cite{Abadi2016Deep}, \cite{geyer2017differentially}, and \cite{jiang2019lightweight}, and collaborative training, e.g., \cite{Shokri2015Privacy} and~\cite{hitaj2017deep}. However, the adoption of these approaches sacrifices the performance, i.e., model accuracy. They also require significant computation on participating mobile devices. Thus, the tradeoff between privacy guarantee and system performance has to be well balanced when implementing the FL system. 

\item \textit{Unlabeled data:} It is important to note that the approaches reviewed in the survey are proposed for supervised learning tasks. This means that the approaches assume that labels exist for all of the data in the federated network. However, in practice, the data generated in the network may be unlabeled or mislabeled \cite{gu2019reaching}. This poses a big challenge to the server to find participants with appropriate data for model training. Tackling this challenge may require the challenges of
scalability, heterogeneity, and privacy in the FL systems to be addressed. One possible solution is to enable mobile devices to construct their labeled data by learning the ``labeled data'' from each other. Emerging studies have also considered the use of semi-supervised learning inspired techniques \cite{albaseer2020exploiting}.

\item \textit{Interference among mobile devices:} The existing resource allocation approaches, e.g.,  \cite{nishio2018client} and \cite{anh2018efficient}, address the participant selection based on the resource states of their mobile devices. In fact, these mobile devices may be geographically close to each other, i.e., in the same cell. This introduces an interference issue when they update local models to the server. As such, channel allocation policies may need to be combined with the resource allocation approaches to address the interference issue. While studies in \cite{zhu2018low}, \cite{amiri2019federated}, and \cite{yang2018federated} consider multi-access schemes and over-the-air computation, it remains to be seen if such approaches are scalable, i.e., able to support a large federation of many participants. To this end, data driven learning based solutions, e.g., federated DRL, can be considered to model the dynamic environment of mobile edge networks and make optimized decisions.

\item \textit{Communication security:} The privacy and security threats studied in Section \ref{sec:securityissues} revolve mainly around data-related compromises, e.g., data and model poisoning. Due to the exposed nature of the wireless medium, FL is also vulnerable to communication security issues such as Distributed Denial-of-Service (DoS) \cite{lau2000distributed} and jamming attacks \cite{xu2006jamming}. In particular, for jamming attacks, an attacker can transmit radio frequency jamming signals with high power to disrupt or cause interference to the communications between the mobile devices and the server. Such an attack can cause errors to the model uploads/downloads and consequently degrade the performance, i.e., accuracy, of the FL systems. Anti-jamming schemes \cite{strasser2008jamming} such as frequency hopping, e.g., sending one more copy of the model update over different frequencies, can be adopted to address the issue.

\item \textit{Asynchronous FL:} In synchronous FL, each training round only progresses as quickly as the slowest device, i.e., the FL system is susceptible to the straggler effect. As such, asynchronous FL has been proposed as a solution in \cite{sprague2018asynchronous} and \cite{xie2019asynchronous}. In addition, asynchronous FL also allows participants to join the FL training halfway even while a training round is in progress. This is more reflective of practical FL settings and can be an important contributing factor towards ensuring the scalability of FL. However, synchronous FL remains to be the most common approach used due to convergence guarantees \cite{bonawitz2019towards}. Given the many advantages of asynchronous FL, new asynchronous algorithms should be explored. In particular, for future proposed algorithms, the convergence guarantee in a non-IID setting for non-convex loss functions need to be considered. An approach to be considered is the possibile inclusion of \textit{backup} workers following the studies of \cite{chen2016revisiting}.

\item \textit{Comparisons with other distributed learning methods:} Following the increased scrutiny on data privacy, there has been a growing effort on developing new privacy preserving distributed learning algorithms. One study proposes \textit{split learning} \cite{vepakomma2018split}, which also enables collaborative ML without requiring the exchange of raw data with an external server. In split learning, each participant first trains the neural network up to a cut layer. Then, the outputs from training are transmitted to an external server that completes the other layers of training. The resultant gradients are then back propagated up to the cut layer, and eventually returned to the participants to complete the local training. In contrast, FL typically involves the communication of full model parameters. The authors in \cite{singh2019detailed} conduct an empirical comparison between the communication efficiencies of split learning and FL. The simulation results show that split learning performs well when the model size involved is larger, or when there are more participants involved, since the participants do not have to transmit the weights to an aggregating server. However, FL is much easier to implement since the participants and FL server are running the same global model, i.e., the FL server is just in charge of aggregation and thus FL can work with one of the participants serving as the master node. As such, more research efforts can be directed towards guiding system administrators to make an informed decision as to which scenario warrants the use of either learning methods.

\item \textit{Further studies on learning convergence:} One of the essential considerations of FL is the
convergence of the algorithm. FL finds weights to minimize the global model aggregation. This is actually a distributed optimization problem, and the convergence is not always guaranteed. Theoretical analysis and evaluations
on the convergence bounds of the gradient descent based
FL for convex and non-convex loss functions are important research directions. While existing studies have covered this topic, many of the guarantees are limited to restrictions, e.g., convexity of the loss function.

\item \textit{Usage of tools to quantify statistical heterogeneity:} Mobile devices typically generate and collect data in a non-IID manner across the network. Moreover, the number of data samples among the mobile devices may vary significantly. To improve the convergence of FL algorithm, the statistical heterogeneity of the data needs to be quantified. Recent works, e.g., \cite{eliazar2010measuring}, have developed tools for quantifying statistical heterogeneity through metrics such as local dissimilarity. However, these metrics cannot be easily calculated over the federated network before training begins. The importance of these metrics motivates future directions such as the development of efficient algorithms to quickly determine the level of heterogeneity in federated networks.

\item \textit{Combined algorithms for communication reduction:} Currently, there are three common techniques of communication reduction in FL as discussed in Section \ref{sec: communication}. It is important to study how these techniques can be combined with each other to improve the performance further. For example, the model compression technique can be combined with the edge server-assisted FL. The combination is able to significantly reduce the size of model updates, as well as the instances of communication with the FL server. However, the feasibility of this combination has not been explored. In addition, the tradeoff between accuracy and communication overhead for the combination technique needs to be further evaluated. In particular, for simulation results we discuss in Section \ref{sec: communication}, the accuracy-communication cost reduction tradeoff is difficult to manage since it varies for different settings, e.g., data distribution, quantity, number of edge servers, and number of participants.

\item \textit{Cooperative mobile crowd ML:} In the existing approaches, mobile devices need to communicate with the server directly and this may increase the energy consumption. In fact, mobile devices nearby can be grouped in a cluster, and the model downloading/uploading between the server and the mobile devices can be facilitated by a ``cluster head'' that serves as a relay node \cite{leu2014energy}. The model exchange between the mobile devices and the cluster head can then be done in Device-to-Device (D2D) connections. Such a model can improve the energy efficiency significantly. Efficient coordination schemes for the cluster head can thus be designed to further improve the energy efficiency of a FL system. 

\item \textit{Applications of FL:} Given the advantages of guaranteeing data privacy, FL has an increasingly important role to play in many
applications, e.g., healthcare, finance and transport systems. For most current studies on FL applications, the focus mainly lies in the federated training of the learning model, with the implementation challenges neglected.
For future studies on the applications of FL, besides
a need to consider the aforementioned issues in the survey,
i.e., communication costs, resource allocation, and privacy
and security, there is also a need to consider the specific issues related to the system model in which FL will be adopted in. For example, for delay critical applications, e.g, in vehicular networks, there will be more emphasis on training efficiency and less on energy expense.

\end{itemize}
%====================================================================
%====================================================================
\section{Conclusion}
\label{sec:conclusions}
This paper has presented a tutorial of FL and a comprehensive survey on the issues regarding FL implementation. Firstly, we begin with an introduction to the motivation for MEC, and how FL can serve as an enabling technology for collaborative model training at mobile edge networks. Then, we describe the fundamentals of DNN model training, FL, and system design towards FL at scale. Afterwards, we provide detailed reviews, analyses, and comparisons of approaches for emerging implementation challenges in FL. The issues include communication cost, resource allocation, data privacy and data security. Furthermore, we also discuss the implementation of FL for privacy-preserving mobile edge network optimization. Finally, we discuss challenges and future research directions.
%====================================================================
%====================================================================

%\begin{thebibliography}{100}
\bibliographystyle{IEEEtran}
\bibliography{FederatedLearning}

\end{document}